\documentclass{aa}  

\usepackage{graphicx}
\usepackage{txfonts}
\usepackage{xcolor}
\newcommand{\astext}[1]{{\textcolor{black}{#1}}}
\newcommand{\gnstext}[1]{{\textcolor{black}{#1}}}
\newcommand{\newgnstext}[1]{{\textcolor{black}{#1}}}
\newcommand{\newastext}[1]{{\textcolor{black}{#1}}}

\begin{document}

   \title{A new mass estimate method with hydrodynamical atmospheres \\ for very massive WNh stars}


\author{{Gautham N. Sabhahit\inst{\ref{AOP}}}
    \and 
       {Jorick S. Vink\inst{\ref{AOP}}}
    \and
      {Andreas\,A.\,C. Sander\inst{\ref{ARI}}}
    \and
      {Matheus Bernini-Peron\inst{\ref{ARI}}}
    \and
      {Paul\,A. Crowther\inst{\ref{UoS}}}
    \and
      {Roel R. Lefever\inst{\ref{ARI}}}
    \and
      {Tomer Shenar\inst{\ref{TAU}}}
    }

\institute{
   {Armagh Observatory and Planetarium, College Hill, Armagh BT61 9DG, N. Ireland\label{AOP}}
   \and
   {Zentrum f{\"u}r Astronomie der Universit{\"a}t Heidelberg, Astronomisches Rechen-Institut, M{\"o}nchhofstr. 12-14, 69120 Heidelberg, Germany \label{ARI}}
   \and
   {Department of Physics \& Astronomy, University of Sheffield, Hicks Building, Hounsfield Road, Sheffield S3 7RH, United Kingdom\label{UoS}}
   \and
   {The School of Physics and Astronomy, Tel Aviv University, Tel Aviv 6997801, Israel\label{TAU}}\\
              \email{gauthamns96@gmail.com}
             }


 
  \abstract
   {Very massive stars with masses over $100\,M_\odot$ are key objects in the Universe for our understanding of chemical and energetic feedback in the Universe, but their evolution and fate are almost entirely determined by their wind mass loss. 
   Here, we aim to determine the mass of the most massive star known in the Local Group R136a1. 
   To this end, we computed the first hydrodynamically consistent non-local thermodynamical equilibrium atmosphere models for R136a1 {(WN5h),} as well as the binary system R144 {(WN5/6h+WN6/7h)} in 
   the Tarantula Nebula. Using the Potsdam Wolf-Rayet code, we were able to simultaneously empirically derive {\it and} theoretically predict their mass-loss rates and wind velocities. By fitting synthetic spectra derived from these models to multi-wavelength observations, we constrained the stellar and wind properties of R144 and R136a1. We first determined the clumping stratification required by our hydro-models to fit the spectra of R144, using the available dynamical mass estimates for the two components. We then utilised this clumping stratification in hydrodynamic models of R136a1 and estimated a mass of $M_\mathrm{Hydro}$ of $233\,M_\odot$. Remarkably, the estimated mass is close to and fully consistent with chemical homogeneous mass relations. This present-day mass of $233\,M_\odot$ provides a lower limit to the initial stellar mass, which could be far higher due to previous wind mass loss.}


   \maketitle
%

\section{Introduction}
\label{sec: Introduction}

Very massive stars (VMSs) are among the most luminous and powerful objects in the Universe. Defined as stars born with initial masses exceeding $100\,M_\odot$, a single VMS can dominate the mechanical and chemical feedback within a young cluster \citep[e.g.][]{Doran2013, Crowther2016,Roy2020,Vink23,Tapia24,Rivera24,Rossi24}. With luminosities exceeding a million times that of the Sun, these stars drive powerful radiation-driven stellar winds, which produce the strong emission lines characteristic of their Wolf-Rayet (WR) spectral type \citep{Graf2011,Vink2011}. The hydrogen-rich Wolf-Rayet stars of the nitrogen (N) sequence (WNh stars), found in young ($\sim 1-2$ Myr) massive clusters, are the best candidates for very massive stars \citep{Massey1998, Dekoter1998, Martins2008,  Wofford2014,Smith23,Mestric2023,Upadhyaya2024}.  Understanding the wind properties of WNh stars is crucial not only for tracing their evolution and ultimate fates, but also for understanding how they shape their surrounding medium through feedback processes.

The most fundamental property of a star is its stellar mass, $M_\star$. For VMSs, there can be a significant difference between their current-day mass and initial mass due to strong mass loss driven by stellar winds, making the back-tracing to infer the initial mass almost impossible and the stellar upper mass-limit unrestricted \citep{Vink2018}. One approach to breaking this degeneracy is to perform stellar spectroscopy to determine the surface hydrogen (H) abundance, $X$. {With the total luminosity and surface H abundance, a mass called the homogeneous mass can be estimated by assuming a fully chemically homogeneous star, that is, with a constant chemical composition throughout \citep{Graf2011}. For VMSs, the chemical homogeneity assumption holds quite well due to their large convective core-to-total mass ratios \citep{Yusof2013, Kohler2015}} and the degeneracy between initial and current-day mass can be potentially resolved \citep{Sabhahit2022, Higgins2022}. However, a key component in this quest is to determine the present-day mass.

For canonical massive OB stars this is generally performed via the determination of surface gravity $\log g$ in photospheric lines, but for WR stars ({including the WNh type)}, this is not possible as the lines are in emission. \gnstext{One way around this would be by comparing to detailed evolutionary models {\citep[for e.g.,][]{Crowther2010, Hamann2019, Brands22}}. However, such evolutionary masses can be unreliable due to uncertain structure physics and mass-loss rates in evolution codes.} 

Another way would be to utilise sophisticated non-local thermodynamical equilibrium (nLTE) stellar atmosphere modelling techniques in co-moving frame (CMF) that include the wind hydrodynamics. Several stellar atmosphere codes exist in the literature that use the CMF radiative transfer approach, including \textsc{wmbasic} \citep{Pauldrach2001}, \textsc{cmfgen} \citep{Hillier1990, Hillier1998}, \textsc{fastwind} \citep{Puls2005}, and PoWR \citep{Hamann85, Hamann86, Grafener2002, HG2003}. Over the years, these codes have been updated to solve the wind hydrodynamics consistently. This has allowed them not only to produce synthetic spectra, but also to simultaneously predict both the wind stratification and the mass-loss rate \citep{GH2005, Sander2017, Gormaz_Matamala2021, Bjorklund2021}. Solving the wind hydrodynamics could potentially deliver a mass estimate as stellar mass enters the hydrodynamic equation of motion affecting mass loss and wind velocities. 

However, stellar winds are clumped {\citep{Eversberg1988, Hillier1991, Hamann1998, Davies2007}}, with not only an uncertain clumping factor (which could be determined from spectral fitting) but also an uncertain stratification and clumping origin. Traditionally, hot-star winds are considered to be intrinsically unstable \citep{LS1970,CAK1975} due to a strong instability, called the line-de-shadowing instability \citep[][]{Owocki2015}, leading to clumping in the wind when the velocities become sufficiently large to produce shocked structures in both density and velocity. However, in more recent years, the sub-photospheric origin of clumping has come to the fore \citep{Cantiello2009,Jiang2015,Schultz2020,Debnath2024}, making the clumping stratification highly uncertain at the present time.

In other words, while hydrodynamical model atmosphere fitting could potentially deliver the stellar mass as an important output, there could be degeneracy with the clumping stratification. However, there is a promising prospect to this as some binary systems comprises of two WNh stars, such as R144, where reliable dynamical mass estimates for both the WNh stars are possible.

{Both the components of R144 and the single object R136a1 are VMSs that are extremely luminous, exhibit similar wind strengths, share similar spectral types (early WNh) and belong to the same nebula, implying comparable abundances.} So, our strategy is to employ the hydrodynamical branch of the non-LTE Potsdam Wolf-Rayet ($\texttt{PoWR}^\textsc{hd}$) code \citep{Sander2017,Sander2020a,Sander2023} to first determine the clumping stratification of both WNh stars in the R144 binary system for which reliable dynamical masses are already available. {Given their comparable wind and stellar properties, we used the above determined clumping stratification for R144 components in our R136a1 hydrodynamical model,} with a flexible volume filling factor obtained from traditional spectral fitting, to estimate the mass of the current record holder in a completely new and independent way. 

The paper is organised as follows. In $\mathrm{Sect.}\,\ref{sec: Observational_data}$, we describe the photometric and spectroscopic data of R144 and R136a1 we use. In $\mathrm{Sect.}\,\ref{sec: stellar_atm_models}$, we detail the basics of the $\texttt{PoWR}$ and the updated hydro-branch. This is followed by a didactic overview of how different input parameters affect the wind properties and the synthetic spectra ($\mathrm{Sect.}\,\ref{sec: Hydro_game}$). In $\mathrm{Sect.}\,\ref{sec: Hydro_model_spectra_fit_R144}$ and $\mathrm{Sect.}\,\ref{sec: Hydro_model_spectra_fit_R136}$, we present our best-fit hydrodynamic models of R144 and R136a1. A brief discussion on the clumping stratification we find is presented in $\mathrm{Sect.}\,\ref{sec: discussion}$ and we conclude in $\mathrm{Sect.}\,\ref{sec: conclusions}$.

\section{Observational data}
\label{sec: Observational_data}

Below, we briefly describe the observational data used in our model comparisons, incorporating both photometric and spectroscopic data across a broad wavelength range.

\subsection{R144: Binary WNh star in the Tarantula region of the LMC}
\label{sec: R144_obs}

Identified as a WN5/6h + WN6/7h double-lined spectroscopic binary system (SB2; \citealt{Sana2013_R144}), R144 (BAT99 118, {HD 38282}) is the visually brightest WR star in the Large Magellanic Cloud (LMC). A detailed orbital analysis by \citet{Tomer2021} recently derived orbital parameters, including dynamical mass estimates for both the primary and secondary components {of $74$ and $69\,M_\odot$, respectively}. {Both components of the R144 system were likely born with initial masses exceeding $100\,M_\odot$, making them individually VMSs \citep{Tomer2021, Higgins2022}. We model the R144 system in this study due to the availability of both dynamical mass estimates and multi-wavelength ultraviolet (UV) + optical + near-infrared (IR) spectroscopic data, making R144 system an ideal test-bed for spectral fitting using hydrodynamical atmosphere modelling over other WNh + WNh systems such as Melnick 34 (Mk34, BAT99 116) studied in \citet{Tehrani2019}, which lacks UV data.} 

{We use observational data, which is largely similar to that detailed in \citet{Tomer2021}.} For photometric data, we used VizieR: BV photometry magnitude from \citet{Parker1992}, R magnitude from \citet{Monet1998}, 2MASS JHK-bands and IRAC magnitudes from \citet{Bonanos2009}, and WISE magnitudes from \citet{Cutri2014}. For optical and near-IR spectra, we use the X-shooter spectroscopy detailed in \citet{Tomer2021}. We also retrieve three far-UV {International Ultraviolet Explorer} (IUE) and {Far Ultraviolet Spectroscopic Explorer} (FUSE) spectra from the Mikulski Archive for Space Telescopes covering spectral range 900-$3000~\text{\AA}$. For additional details, we refer to \citet{Tomer2021}.

\subsection{R136a1}
\label{sec: R136a1_obs}

R136 is a young, massive star cluster located at the centre of the Tarantula (30 Dor) star-forming region in the LMC \citep{Hunter1995, Massey1998, Khorrami2021}. This cluster hosts several VMS with estimated masses exceeding $100\,M_\odot$ \citep{Crowther2010, Crowther2016, Best2020}.

In this work, we model the most luminous of the very massive stars in this cluster, the WN5h star R136a1 (BAT99 108). \gnstext{R136a1 is apparently  a single star and, so far, no spectroscopic evidence exists for this star being in a binary system \citep{Shenar2023}. There is a close visual companion however, catalogued by \citet{Hunter1995} and most recently revisited by \citep{Kalari22}. While it is possible that additional photometric components may still be discovered, the fact remains that these WNh stars show strong emission-line spectra and would not simply resolve into a bunch of lower luminosity O stars that have typical absorption-line-dominated spectra.}

For optical data, we use multi-colour Hubble Space Telescope (HST)/WFC3 photometry in the F336W, F438W, F555W, and F814W bands, provided by \citet{DeMarchi2011}. In the near-IR, we utilise {the Ks-band photometry from the SPHERE instrument on the Very Large Telescope (VLT)} \citep{Khorrami2017}. For spectroscopic data, we use the following observations: far-UV HST-STIS/G140L, blue-optical HST-STIS/G430M, and H$\alpha$ {HST-STIS}/G750M from \citet[][]{Crowther2016}. In the near-IR, we incorporate spatially resolved K-band spectroscopy from VLT/SINFONI, obtained by \citet{Schnurr2009}. For more information, see \citet{Crowther2010, Crowther2016, Best2020}.

\section{Stellar atmosphere models}
\label{sec: stellar_atm_models}

In this section, we give a brief overview of the $\texttt{PoWR}$ atmosphere code. The base and hydro-branch of the code are discussed in $\mathrm{Sect.}\,\ref{sec: Powr_basics}$ and the input parameters are detailed in $\mathrm{Sect.}\,\ref{sec: Input_parameters}$.

\subsection{Basic concepts}
\label{sec: Powr_basics}

We used the non-LTE code $\texttt{PoWR}$ \citep{Grafener2002, HG2003, Sander2015} to compute our model stellar atmospheres. The standard version of the code models a spherically symmetric, expanding non-grey atmosphere with a stationary outflow. The velocity field is originally \textit{prescribed}, typically as a $\beta-$law attached to a solution of the hydrostatic equation. Given the velocity field and an \textit{input} mass-loss rate, the density stratification then follows from the continuity equation of
\begin{equation}
\begin{array}{c@{\qquad}c}
\dot{M} =  4 \pi r^2 \rho(r) \varv(r) = \mathrm{constant.}
\end{array}
\label{eq: continuity}
\end{equation}
Given this basic stratification, the code then iteratively solves the radiative transfer in CMF \citep[based on the concepts of][]{Mihalas1975, Mihalas1976} and a set of statistical equilibrium equations. In total, this results in a system of (non-linear) equations, fully coupled in the spatial and frequency domains.
In addition, the determination of the temperature stratification is also part of the iterative process. It is obtained from a correction scheme based on the Uns{\"o}ld-Lucy method \citep{Unsold1955, Lucy1964}, generalised to expanding, non-grey atmospheres \citep{HG2003}. Alternatively, temperature corrections can also be obtained from electron thermal balance \citep{Kubat1999,Sander2015diss}.
To facilitate the convergence of the overall iteration cycle, the calculation of the non-LTE population numbers includes an Accelerated Lambda Iteration approach \citep[see][]{Hamann85, Hamann86} with a flexible damping depending on the maximum correction to significantly populated levels. The overall iteration cycle continues until the corrections for the population numbers are sufficiently low and a conservation of the flux as well as the total Rosseland continuum optical depth is ensured, at which point the model is considered to be converged. Further details on of the $\texttt{PoWR}$ code can be found in \citet{Grafener2002} and \citet{HG2004}.

From a converged model, a synthetic emergent spectrum can be generated by performing a formal integration in the observer's frame including relevant broadening processes. The synthetic spectrum is then compared to an observed spectrum. If the fit is poor, the input parameters are varied, a new model atmosphere is calculated, and the spectra are compared again. This process continues until a satisfactory fit is obtained, allowing for the empirical determination of key stellar parameters and the wind stratification.

Alternatively to the standard branch with a pre-specified assumption of the velocity field in the wind, the $\texttt{PoWR}^\textsc{hd}$ branch \citep{Sander2017,Sander2020a,Sander2023} allows us to solve the time-independent, stationary hydrodynamic equation of motion,
\begin{equation}
\begin{split}
\varv\dfrac{\mathrm{d}\varv}{\mathrm{d}r} = - \dfrac{1}{\rho}\dfrac{\mathrm{d}P_\mathrm{tot}}{\mathrm{d}r} - \dfrac{GM_\star}{r^2},
\end{split}
\label{eq: hydro_eq}
\end{equation}
as a part of the overall iteration. The expression in Eq.\,\eqref{eq: hydro_eq} describes the force balance in the 1D stellar atmosphere where the advective acceleration term on the left depends on the difference between the total pressure gradient and Newtonian gravity, where $\rho$ is the density, and $P_\mathrm{tot}$ is the sum of gas, turbulent, and radiation pressure.
When solving Eq.\,\eqref{eq: hydro_eq}, the obtained velocity field is consistent with the different forces in the atmosphere, both in the quasi-hydrostatic layers near the photosphere and in the wind, where terminal velocities ($\varv_\infty$) of several thousands of km\,s$^{-1}$ can be realised. The mass-loss rate $\dot{M}$ does not appear explicitly in Eq.\,\eqref{eq: hydro_eq}, but is implicitly fixed by a conservation condition, typically the pre-specified total optical continuum optical depth. The coupling of the non-LTE atmosphere calculations with the solution of stationary hydrodynamics in these `next-gen' $\texttt{PoWR}^\textsc{hd}$ atmosphere models therefore enable to simultaneously predict the wind stratification as well as the emergent spectra from a given set of input stellar parameters.

Although the mass-loss rate,  $\dot{M}$, and the velocity field, $\varv(r),$ are no longer prescribed inputs in hydrodynamic models, Eq.\,\eqref{eq: continuity} still holds, connecting $\dot{M}$, $\varv(r)$, and the density stratification, $\rho(r),$ and it is used to update $\rho(r)$ every time the $\dot{M}$ and $\varv(r)$ values are updated. 
To gain a new $\varv(r)$, Eq.\,\eqref{eq: hydro_eq} is rewritten as a function of the velocity gradient and then integrated inwards and outwards from its critical point, $r_\mathrm{crit}$, where the denominator of the rewritten hydrodynamic equation of motion goes to zero \citep[for numerical details regarding the implementation of the hydrodynamics in $\texttt{PoWR}$, see][]{Sander2017}. As the radiative acceleration, $a_\text{rad}$, is given as a function of radius from the CMF calculations, $r_\mathrm{crit}$ is also the location where the flow velocity is $\varv(r_\mathrm{crit}) = (\varv_\mathrm{sound}^2 + \varv_\mathrm{turb}^2)^{1/2}$, where $\varv_\mathrm{sound}$ and $\varv_\mathrm{turb}$ are the isothermal sound speed and the turbulent velocity, respectively. An updated velocity and density stratification can change the resulting radiation field, which, in turn, affects the population numbers, and the iteration continues until a satisfactory convergence is obtained.

\subsection{Input parameters}
\label{sec: Input_parameters}

The input parameters for $\texttt{PoWR}^\textsc{hd}$ models are the same as for the base version of the code, except that values for the mass-loss rate or the terminal wind velocity are only treated as starting values, which can also be taken from an old $\texttt{PoWR}$ or $\texttt{PoWR}^\textsc{hd}$ model. Below, we briefly summarise the inputs used in our modelling. 

The stellar luminosity, $L_\star$, the effective temperature, $T_\mathrm{\star}$, at the inner boundary defined at $R_\star$, are connected by the Stefan-Boltzmann equation:
\begin{equation}
\begin{array}{c@{\qquad}c}
L_\star = 4 \pi  \sigma_\mathrm{SB} R_\star^2 T_\mathrm{\star}^4. 
\end{array}
\label{eq: Stefan_Boltzmann}
\end{equation}
Any two of the aforementioned three quantities need to be given. 

The inner boundary is defined at a maximum Rosseland continuum optical depth $\tau_\mathrm{Ross, cont}$, that is, $T_\star = T_\mathrm{eff}(r=R_\star)$ is defined where  $\tau_\mathrm{Ross, cont}$ is equal to a specified value. Models presented in this work use $\tau_\mathrm{Ross, cont} = 5$. Given the high luminosity-to-mass ratios and the temperature regimes of the VMS objects modelled here, we have not included any deeper layers in the atmosphere, as the models could run into a regime with super-Eddington acceleration across the iron-opacity bump \citep[though there are also opacity-reduction arguments from multi-D simulations, e.g. in][]{Shaviv1998, Jiang2015, Schultz2020, Debnath2024} As we  discuss in a separate modelling paper (Lefever et al., submitted), the observed spectra of very massive WNh stars cannot be explained by deep wind launching from the (hot) iron-opacity bump;  therefore, we decided to exclude the complicated deeper regions from this work to focus on the wind launching and resulting spectra. 

The outer boundary is set to a fixed multiple of the stellar radius of the model, which is fixed to $1000\,R_\star$ in this work. The stellar mass, $M_\star$, can either be provided explicitly or calculated from $\log g$ (and $R_\star$). Alternatively, the mass can also be estimated from mass-luminosity relations \citep{Langer1989, Graf2011}. The chemical composition is specified as mass fractions, which are kept constant throughout the atmosphere. 

Density inhomogeneities in the wind {\citep{Puls2008, Hamann2008}} are treated using the so-called micro-clumping approach, where clumped regions are assumed to be optically thin, and inter-clump regions are treated as voids. We test various depth-dependent clumping stratifications, including constant clumping, as well as stratifications that increase or decrease with radius \citep[see][for diagrams]{Muijres2011}.

The pressure gradient acceleration term in the hydrodynamic equation (see $\mathrm{Eq.}\,\ref{eq: hydro_eq}$) can be separated into the following terms:
\begin{equation}
\begin{split}
- \dfrac{1}{\rho}\dfrac{\mathrm{d}P_\mathrm{tot}}{\mathrm{d}r} &= a_\mathrm{gas} + a_\mathrm{turb} + a_\mathrm{rad} \\&
= -\dfrac{1}{\rho}\dfrac{\mathrm{d}(\rho \varv_\mathrm{sound}^2)}{\mathrm{d}r} -\dfrac{1}{\rho}\dfrac{\mathrm{d}(\rho \varv_\mathrm{turb}^2)}{\mathrm{d}r} + \dfrac{\varkappa_\mathrm{F}(r)L(r)}{4\pi r^2 c}
\end{split}
\label{eq: pressure_grad}
,\end{equation}
where $\varkappa_\mathrm{F}(r)$ is the flux-weighted mean opacity, $L(r)$ is the local luminosity. All quantities in the above equation are locally defined, that is, they vary as a function of radius $r$. The quantity $\varv_\mathrm{turb}$ can be specified in our models and is kept constant across our model atmospheres.

The local luminosity, $L(r)$, is not exactly constant in the atmosphere, as energy is required for lifting the material out of the potential well of the star and accelerating it up to the terminal velocity. However, this effect is minimal. For instance, even for the most optically thick winds presented in this work, the difference between the inner and outer boundary luminosities is below $5\%$. This effect can therefore be safely neglected, that is, we can assume that $L(r) = 4 \pi  \sigma_\mathrm{SB} r^2 T_\mathrm{eff}(r)^4 \approx L_\star$.

\begin{figure*}
    \includegraphics[width = \textwidth]{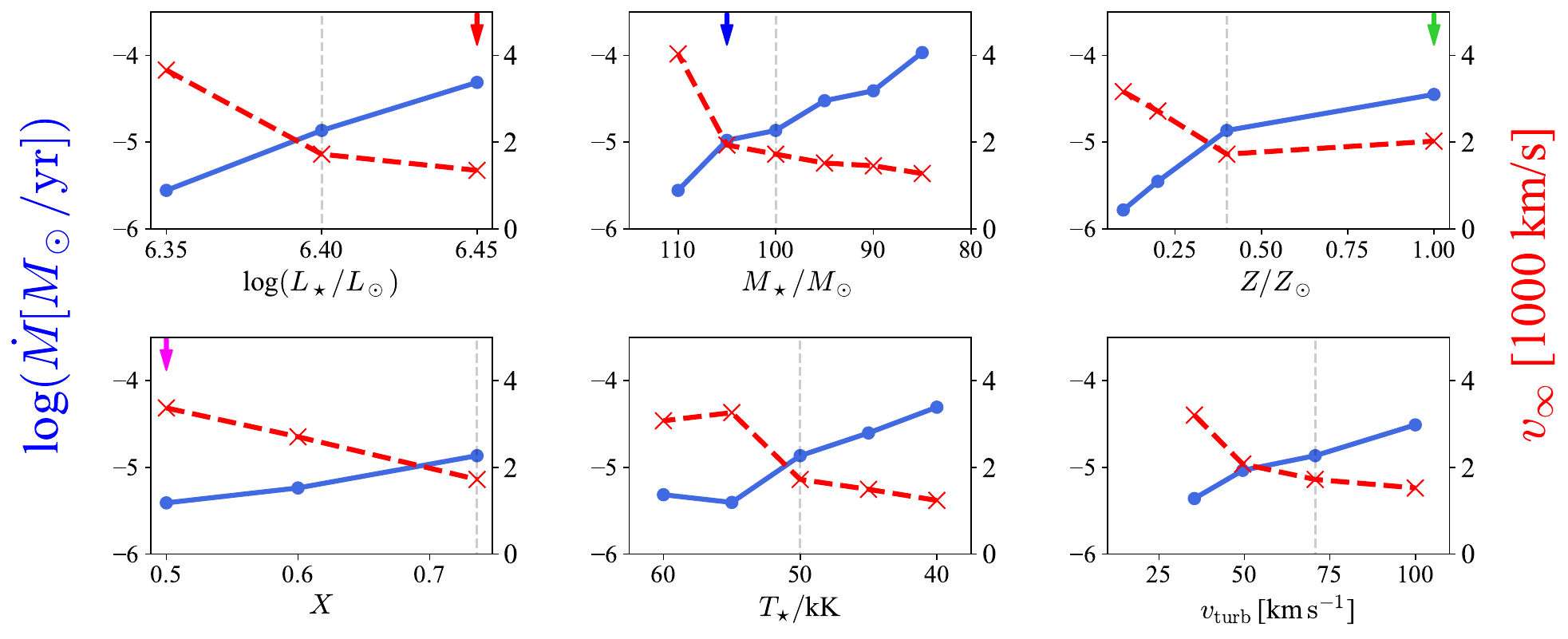}
    \caption{Predicted mass-loss rates (blue) and terminal velocities (red) from hydrodynamic atmosphere models. The six sub-plots illustrate the effect of individually varying $L_\star$, $M_\star$, $Z$, $X$, $T_\star$, and $\varv_\mathrm{turb}$ relative to the `base model'. The `base model' is indicated by a grey dashed line. A select few models from this grid (marked with coloured arrows) are further investigated in terms of their radiative acceleration and temperature stratification in $\mathrm{Fig.}\,\ref{fig: Gamma_rad_plot}$.}
    \label{fig: hydro_game}
\end{figure*}

\begin{figure*}
    \includegraphics[width = \textwidth]{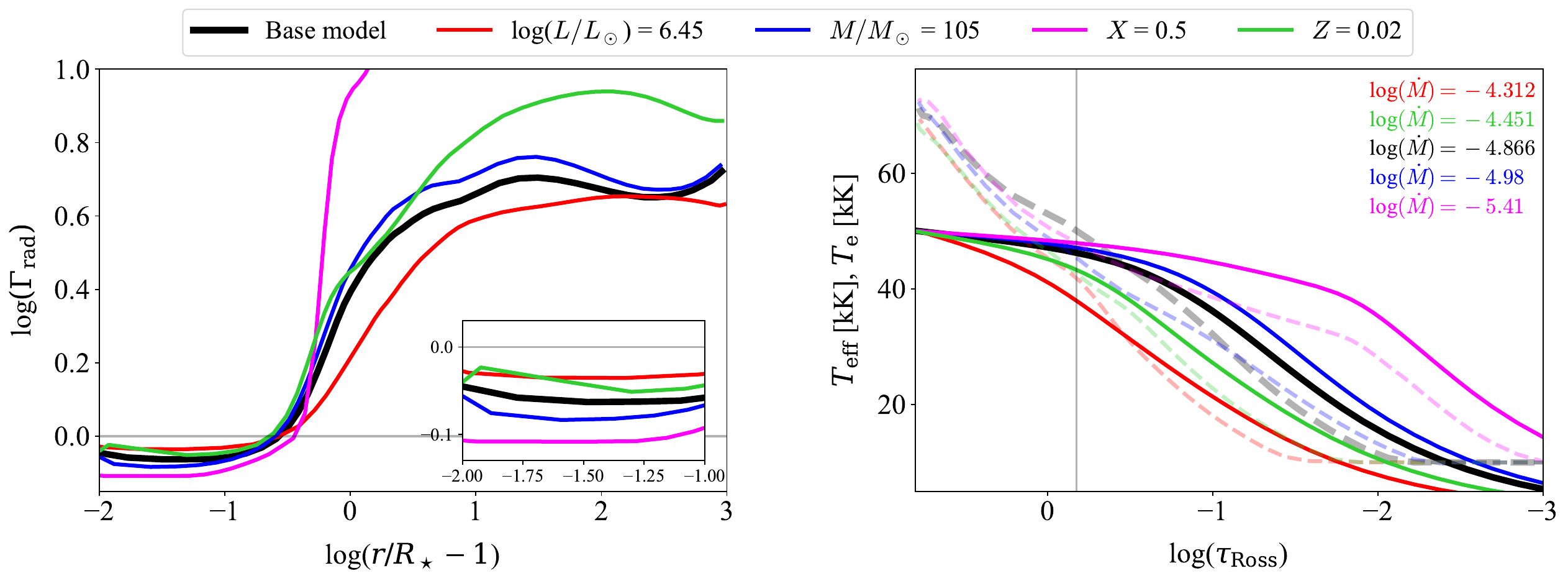}
    \caption{\textit{Left:} Radiative acceleration normalised to gravity. \textit{Right:} Effective temperature (solid) and electron temperature (dashed) stratification for specific models from $\mathrm{Fig.}\,\ref{fig: hydro_game}$. The grey horizontal line in the left panel is the radiative Eddington limit, $\Gamma_\mathrm{rad} = 1$. The inset plot in the left panel shows the zoom-in of the sub-critical region. The grey vertical line in the right panel is marked at $\tau_\mathrm{Ross}=2/3$. In both panels, the black solid line represents the base model, while the coloured solid lines indicate models where each of the following parameter is varied from the `base model': $L$ (red), $M$ (blue), $X$ (magenta), and $Z$ (green). }
    \label{fig: Gamma_rad_plot}
\end{figure*}

\section{The `game' of hydrodynamic spectral modelling }
\label{sec: Hydro_game}

Before presenting hydrodynamic models for the R144 binary stars and R136a1, we would like to first explore how hydrodynamic atmosphere models change the approach to quantitative spectroscopy. When performing spectral analysis with non-hydrodynamic models underlying the modelling, the mass-loss rate and wind stratification are inputs and are entirely decoupled from other input parameters. Whether the prescribed wind can actually drive and sustain the input mass-loss rate via the different forces in the wind is not taken into consideration.

In hydrodynamic models, mass-loss rate and velocity field are no longer direct inputs. They are outputs from the solution to the hydrodynamic equation and can be affected by the other input parameters such as luminosity, surface temperatures, chemical abundances, and so on. For example, increasing the luminosity increases the radiative acceleration in the sub-critical region. The wind becomes denser and slower and drives a higher mass-loss rate.

Using hydrodynamic models also means that while performing spectral analysis, the mass-loss rate and wind stratification need to be controlled by other parameters to fit wind-sensitive lines. In the above example, the changes in the wind (higher $\dot{M}$ and slower velocities) are reflected in the spectra, where recombination lines become even stronger in emission. Even changes to the steepness of the velocity field, which can be captured by an effective $\beta$ in our hydrodynamic models, can also influence the emergent synthetic spectra \citep[e.g. see the tests done by][]{Lefever2023}.

This is a simple case, but the complexity can quickly escalate. For instance, $\texttt{PoWR}^\textsc{hd}$ models use the effective temperature, $T_\star$, at the inner boundary as an input, rather than the effective temperature at $\tau_\mathrm{Ross} = 2/3$. This requires adjusting $T_\star$ to achieve the desired $T_\mathrm{eff}(\tau_\mathrm{Ross} = 2/3)$ value, ensuring that temperature-sensitive lines are satisfactorily fit. However, changing $T_\star$ will also affect the mass-loss rate, the wind velocities, and the ionisation balance in the wind. In fact, the above example of increasing $L$ also affects the relation between $T_\star$ and $T_\mathrm{eff}(\tau_\mathrm{Ross} = 2/3)$. Given these complexities, we first provide a didactic overview of the impact of each individual stellar parameter on the emergent synthetic spectra. 

Spectral modelling using hydrodynamic models can be approached in two steps. First, we examine the effect of individual parameters on the predicted wind properties, mainly the mass-loss rate and wind velocities ($\mathrm{Sect.}\,\ref{sec: stellar_param_mass_loss}$ and $\ref{sec: clumping_effect}$). Afterwards, we check how these changes in mass loss and velocities impact the synthetic emission-line measure ($\mathrm{Sect.}\,\ref{sec: spectra_connection}$). 

\subsection{Effect of stellar parameters on the predicted mass loss}
\label{sec: stellar_param_mass_loss}

To understand the first step, we start with a `base model' with a fixed set of inputs: log$(L_\star/L_\odot) = 6.4,\, T_\star = 50\,\mathrm{kK},\, M_\star/M_\odot = 100, X = 0.736,$ total metal mass fraction, $Z = 0.008$, $\varv_\mathrm{turb} = 70.71\,\mathrm{km\,s}^{-1}$, and a clumping stratification which increases from no clumping to $D_\mathrm{cl}=25$ at the outer boundary using the Hillier prescription \citep{Hillier2003} with characteristic velocity $\varv_\mathrm{cl}=100\,\mathrm{km\,s}^{-1}$.  

From this `base model', we vary one input parameter at a time while keeping all other inputs fixed to isolate the effect of that parameter. In $\mathrm{Fig.}\,\ref{fig: hydro_game}$, we show the hydro-predicted mass-loss rates and the terminal velocities obtained by individually varying $L_\star$, $M_\star$, $Z$, $X$, $T_\star$ and $\varv_\mathrm{turb}$. The complete list of input parameters used in this section, along with the predicted wind properties, can be found in $\mathrm{Appendix}\,\ref{Appendix: powrHD_model_properties}$.

The effect of each varied parameter can be understood by examining how it affects the different forces in the wind. In $\mathrm{Fig.}\,\ref{fig: Gamma_rad_plot}$ (left sub-panel), we show the radial variation of the radiative acceleration for selected models from $\mathrm{Fig.}\,\ref{fig: hydro_game}$. The radiative accelerations are normalised to gravity, which gives the radiative Eddington parameter (see $\mathrm{Eq.}\,\ref{eq: pressure_grad}$):
\begin{equation}
\begin{array}{c@{\qquad}c}
\Gamma_\mathrm{rad} = \dfrac{a_\mathrm{rad}}{g_\mathrm{grav}} = \dfrac{\varkappa_{F}L}{4\pi Gc M_\star}
\end{array}
\label{eq: rad_acceleration}
\end{equation}
The flux-weighted mean opacity $\varkappa_\mathrm{F}$ is obtained by a brute-force CMF radiative transfer solution in our models. The $\Gamma_\mathrm{rad}$ term consists of contributions from both lines as well as continuum processes including the electron scattering ($\sigma_\mathrm{e}$), and free-free and bound-free transitions. By considering only the contribution of electron scattering, one can define the (classical) electron scattering Eddington parameter as:
\begin{equation} \Gamma_\mathrm{e} = \dfrac{\sigma_\mathrm{e}L}{4\pi Gc M_\star},
\label{eq: thom_acceleration}
\end{equation}
The advantage of defining $\Gamma_\mathrm{e}$ is that, unlike the flux-weighted mean opacity, the electron scattering opacity in our hot atmosphere models -- hot enough for H and helium (He) to be ionised -- remains approximately constant with radius. The electron scattering $\Gamma_\mathrm{e}$ therefore provides a depth-independent measure of the proximity of the model to the Eddington limit. 

\begin{figure*}
    \includegraphics[width = \textwidth]{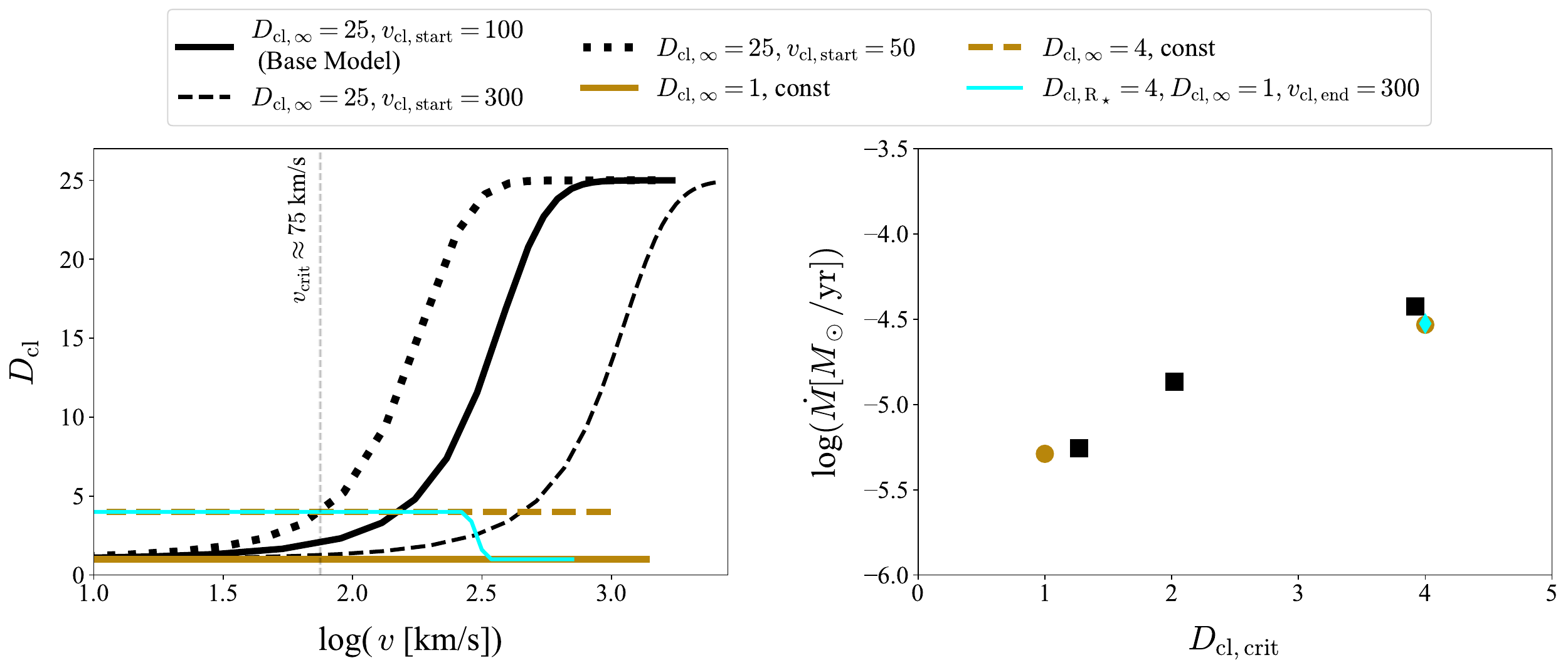}
    \caption{Different stratifications of micro-clumping tested in our hydrodynamic atmosphere models. \textit{Left:} The black solid, dashed, and dotted lines represent clumping stratifications that increase outward, while the cyan line depicts clumping that decreases outward. The brown solid, dashed, and dotted lines represent constant clumping throughout the atmosphere. The grey dashed vertical line indicates the velocity at the critical radius which is roughly $v_\mathrm{crit}\approx75\,\mathrm{km\,s}^{-1}$ for all the models tested here. \textit{Right:} The predicted mass-loss rate as a function of the clumping value at the critical point.}
    \label{fig: denscon}
\end{figure*}

The location where $\Gamma_\mathrm{rad}$ crosses unity roughly coincides with the critical point defined previously. In general, if the total pressure gradient increases (or decreases) in the  sub-critical, inner quasi-hydrostatic region, the mass-loss rate increases (or decreases). If the total pressure gradient increases (or decreases) in the super-critical, outer wind region, then the terminal velocities increase (or decrease) as a result \citep{Lamers1999, Vink99, Sander2017, Sander2020a, Sander2023}. 

A note of caution is necessary here as predicting a priori how $\dot{M}$ and $\varv_\infty$ will vary with different stellar parameters is not trivial. This complexity arises from the non-linear, frequency- and spatially coupled nature of the equations being solved, which means that radiative acceleration as a function of radius cannot be determined locally and will depend on the conditions elsewhere in the atmosphere. Therefore, the trends identified here should  not be generalised beyond the tested ranges. In particular, the terminal velocities are highly sensitive to conditions in the outer wind and can show non-monotonic behaviour in our predicted values \citep[see for example Fig.\,14 in][]{Sander2020b}.
The effects of each of the six parameters on the predicted wind properties can be understood as follows:

\vspace{0.5 cm}
\begin{enumerate}
    \item Luminosity, $L_\star$: As the luminosity increases, the radiative acceleration in the inner wind is higher, and the model drives a denser wind with a higher mass-loss rate. \astext{As  to first order $a_\text{rad} \propto \rho^{-1}$, 
    the increased $\dot{M}$ usually reduces the acceleration in the outer part and $\varv_\infty$ is lower.}
    \item Stellar mass, $M_\star$: Mass enters the force balance equation through the gravity term.  With higher mass, material must be lifted from a deeper gravitational potential well.
    Consequently, the $\Gamma_\mathrm{rad}$ parameter in the sub-critical region decreases. The mass-loss rate therefore decreases as the mass increases. In turn, $\varv_\infty$ weakly increases with increasing mass.
    \item Metallicity, $Z$: As the metal content in the atmosphere increases, line driving in the inner wind becomes stronger, leading to a higher mass-loss rate. The presence of more metals also increases the line acceleration in the outer wind, primarily from elements like C, N, O, and Ar, which tends to increase the terminal velocity. However, the higher mass loss leads to a denser wind, which can, in turn, reduce the terminal velocity. This interaction can produce non-monotonic effects on $\varv_\infty$, which may also arise in other cases discussed here, but is particularly prevalent when varying $Z$ 
    \item H mass fraction, $X$: As the H mass fraction decreases, the electron number density reduces. The contribution from the electron scattering to the total radiative acceleration decreases and the mass-loss rate decreases.
    In the models presented here, $\varv_\infty$ increases with lower $X$.
    \item Temperature, $T_\star$: The effect of temperature on mass loss is complex and can vary across different parameter regimes due to a combination of factors. As temperature decreases, sudden ionization changes may occur in the wind, or the peak of the spectral energy distribution (SED) may shift away from the UV, creating a mismatch between the location of strong wind-driving lines and the SED peak. Additionally, lower temperatures mean larger radii at fixed luminosity, allowing for material to escape the potential well more easily.
    Consequently, non-monotonic trends in both $\dot{M}$ and $\varv_\infty$ can be predicted with changing temperature. We do not plot the $\Gamma_\mathrm{rad}$ parameter for a model with varying temperature in $\mathrm{Fig.}\,\ref{fig: Gamma_rad_plot}$, as its effect on the radiative forces can be non-monotonic and not straight-forward to interpret.
    \item Turbulent velocity, $\varv_\mathrm{turb}$: We also test models with different values of $\varv_\mathrm{turb}$. To first order, this term does not directly influence the radiative acceleration. However, it affects the total pressure gradient through the turbulence term. A higher $\varv_\mathrm{turb}$ effectively increases the total pressure gradient, resulting in an increase in the mass-loss rate and a decrease in $\varv_\infty$. 
\end{enumerate}

Now that we have a cursory understanding of how different parameters affect mass loss and wind velocities, we present an additional layer of complexity inherent to our models. In the right sub-panel of $\mathrm{Fig.}\,\ref{fig: Gamma_rad_plot}$, we plot the variation of the effective temperature, $T_\mathrm{eff}(r) \sim T_\star/r^2$, and the electron temperature against the logarithm of the Rosseland mean opacity. We also mark the effective temperature at $\tau_\mathrm{Ross} = 2/3$, which is a more physically meaningful quantity compared to the inner boundary temperature.

Across the range tested, we find that as the mass-loss rate increases -- that is, when the winds become denser -- the difference between the effective temperature at $\tau_\mathrm{Ross} = 2/3$ and the inner boundary grows \citep[see also][]{Sander2023}. For example, for $\log(\dot{M}[M_\odot/\mathrm{yr}]) = -5.41$, the difference in temperature is 2 kK, whereas at $\log(\dot{M}[M_\odot/\mathrm{yr}]) = -4.312$, the difference increases to 13 kK! This difference can increase drastically with higher mass-loss rates, leading to the formation of extended sub-photospheric layers. The electron temperature is also altered with ramifications for the ionisation balance in the atmosphere.

This phenomenon is not novel in the literature, as such pseudo-photospheres have been proposed before to explain the cyclical variations observed in S-Dor luminous blue variables \citep[LBVs;][]{Smith2004}. However, for the purposes of this work, the varying difference between the input inner boundary temperature and the more physically meaningful temperature at $\tau_\mathrm{Ross}=2/3$ can have important implications when fitting to observed spectra. This is because the inner boundary temperature needs to be continuously varied to keep $T_\mathrm{eff}(\tau_\mathrm{Ross} = 2/3)$ fixed.

\subsection{Tests on clumping stratification}
\label{sec: clumping_effect}

One parameter that was not considered in the previous sub-section is clumping. Several observational findings (including optical and UV line profile variability, electron scattering wings, and spectro-polarimetric variations) indicate that the winds of massive stars are inhomogeneous \citep[for a review, see][]{Puls2008}. Clumping plays an important role in mass-loss line diagnostics. This is because recombination-based processes, such as H$\alpha$ and \ion{He}{ii} $\lambda$4686, as well as the far-infrared and radio continuum (whose opacities scale with the square of the density), are highly sensitive to the presence of density structures in the wind.

Our hydrodynamic models assume time-independent, stationary winds, meaning time-dependent effects, such as density inhomogeneities and velocity-porosity effects, are not linked to the hydro-solution and are not outputs from these models. However, as we show in $\mathrm{Sect.}\,\ref{sec: Hydro_model_spectra_fit_R144}$, fitting spectra while also ensuring dynamical consistency can potentially provide insights into the clumping stratification within the framework of the applied clumping law.

In our models, we can specify a depth-dependent micro-clumping stratification. In $\mathrm{Fig.}\,\ref{fig: denscon}$, we present the different clumping stratifications tested from the `base model' and the corresponding predicted mass-loss rates. The clumping stratification is plotted as a function of velocity, allowing us to also infer the terminal velocities. All other inputs are the same as the `base model' defined earlier. 

We test the following cases: 1. clumping that remains constant throughout the atmosphere $D_\mathrm{cl, const}$; 2. clumping that increases from roughly unity (no clumping) to a specified maximum micro-clumping factor, $D_\mathrm{cl,\infty}$, using the Hillier prescription with the characteristic onset velocity at $\varv_\mathrm{cl}$; and 3. clumping that decreases from a specified maximum micro-clumping factor, $D_\mathrm{cl,R_\star}$, to no clumping below a specified $\varv_\mathrm{cl, end}$, where the transition is smoothed across three depth points.

\begin{figure*}
    \includegraphics[width = \textwidth]{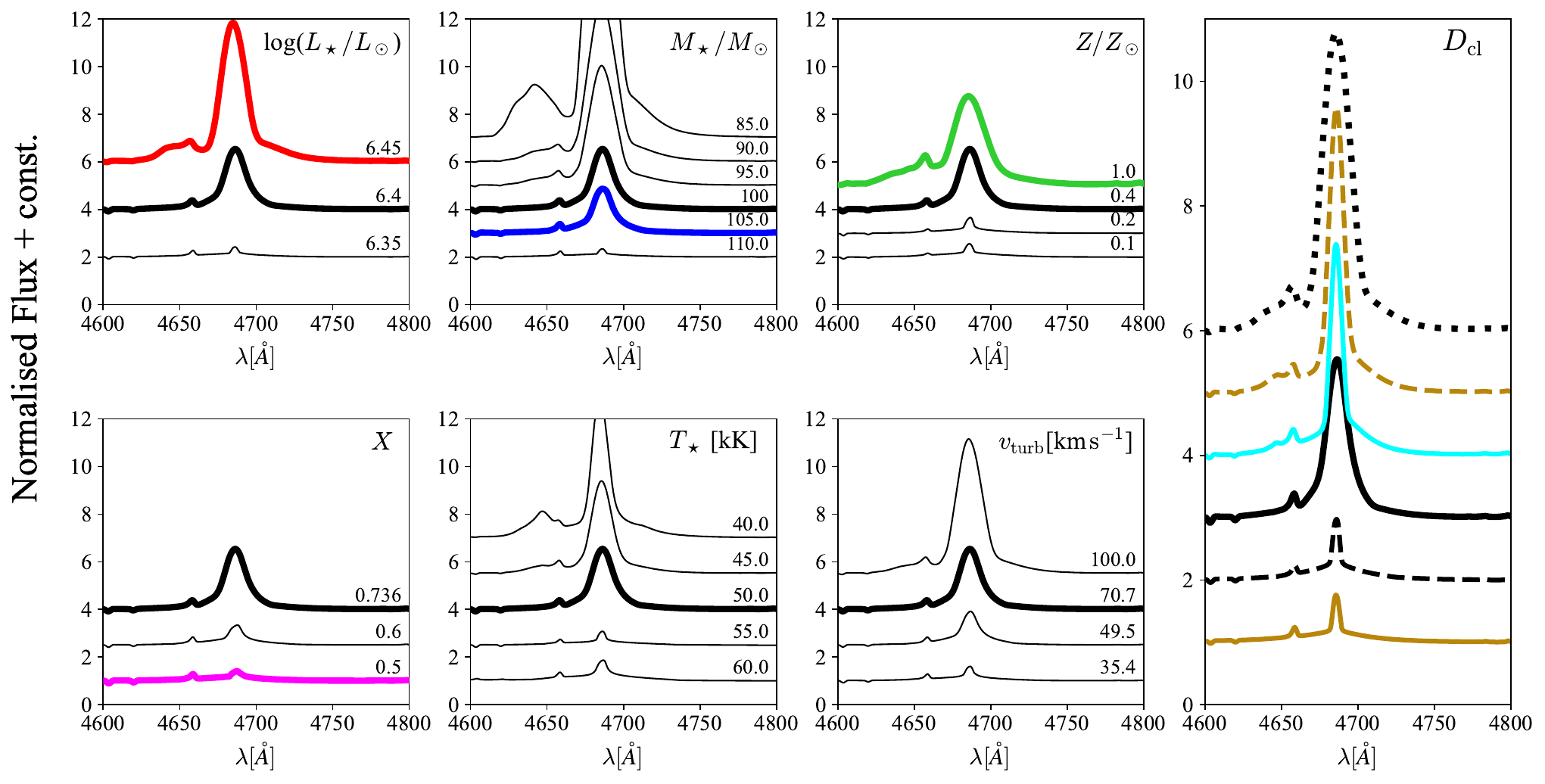}
    \caption{Synthetic \ion{He}{ii} $\lambda$4686 recombination line from hydrodynamic atmosphere models. The black solid line represents the `base model' from Sect. \ref{sec: Hydro_game}. Each sub-plot illustrates the effect of individually varying input parameters (shown in top left corner) on the strength of the synthetic \ion{He}{ii} $\lambda$4686 line. The colour scheme corresponds to $\mathrm{Figs.}\,\ref{fig: hydro_game}$ and $\ref{fig: denscon}$.}
    \label{fig: hydro_game_spectra}
\end{figure*}

We find that the mass-loss rate predicted from our hydrodynamic models depends on the clumping conditions near the critical point \citep[see also,][]{GH2005, Muijres2011}. Models with similar clumping values near the critical point exhibit comparable mass-loss rates. For example, consider the model with no clumping ($D_\mathrm{cl,const} = 1$) and the model with $D_\mathrm{cl,\infty} = 25$ and a clumping onset at $\varv_\mathrm{cl} = 300$ km\,s$^{-1}$. The clumping value of the second model at the critical point is very close to unity, $D_\mathrm{cl,crit} \approx 1$, and the two models have very similar mass-loss rates. The terminal velocity of these two models depends on the clumping in the outer wind. The higher the clumping in the outer wind, the higher the terminal velocity. For example, the no-clumping model has $\varv_\infty = 1368$ km\,s$^{-1}$, while the model with $D_\mathrm{cl,\infty} = 25$ and $\varv_\mathrm{cl} = 300$ km\,s$^{-1}$ has $\varv_\infty = 2527$ km\,s$^{-1}$.

A similar comparison can also be performed for the following two models: constant clumping with $D_\mathrm{cl,const} = 4$ and clumping stratification that starts with $D_\mathrm{cl,R_\star} = 4$ and decreases to no clumping outside the critical point. The two models have similar mass-loss rates, while the model with higher clumping in the outer wind (i.e. the constant clumping model) has a higher terminal velocity.

We also find that the mass-loss rate increases with the clumping value at the critical point. Consider the example of the $D_\mathrm{cl,const} = 1$ and $D_\mathrm{cl,const} = 4$ models. For the temperatures considered here ($T_\star = 50$ kK), the main wind-driving ions are \ion{Fe}{v} and \ion{Fe}{vi}. An ionization change from \ion{Fe}{vi} to \ion{Fe}{v} occurs as the clumping increases, which ultimately affects the predicted mass-loss rate. In such scenarios, predicting terminal velocities becomes non-trivial and will depend on both the inner wind and outer wind clumping conditions.

\subsection{Connecting to the spectra}
\label{sec: spectra_connection}

In the previous two sub-sections, we discuss how different input parameters in our models affect the mass-loss rate and the wind velocities. However, this is only half of the challenge. The other half involves the actual procedure of fitting wind lines from the synthetic spectra to the observed spectra. In $\mathrm{Fig.}\,\ref{fig: hydro_game_spectra}$, we show the normalised \ion{He}{ii} $\lambda$4686 recombination line spectrum obtained for the models presented in $\mathrm{Figs.}\,\ref{fig: hydro_game}$ and $\ref{fig: denscon}$. One immediate trend we notice is that regardless of the input parameter that changes the mass loss, the emission lines get stronger when the mass-loss rate is higher. 

In general, the strength of such recombination lines in emission can be understood through the concept of the transformed mass-loss rate, introduced by \citet{Grafener2013}, which is given by
\begin{equation}
\begin{array}{c@{\qquad}c}
\dot{M}_\mathrm{t} = \dot{M}\times\dfrac{1000\, \mathrm{{km\,s^{-1}}}}{\varv_\infty} \times \Bigg(\dfrac{10^6L_\odot}{L_\star}\Bigg)^{3/4}\times D_\mathrm{cl,\infty}^{1/2}.
\end{array}
\label{eq: transformed_mdot}
\end{equation}
\citet{Best2020} found that the optically-thick line strengths scale well with the transformed mass-loss rate. If the above combination of $\dot{M}$, $\varv_\infty$, $L_\star$, and $D_\mathrm{cl}$ results in a higher $\dot{M}_\mathrm{t}$, the emission line is stronger. Conversely, models with different combinations of $\dot{M}$, $\varv_\infty$, $L_\star$, and $D_\mathrm{cl}$ but similar $\dot{M}_\mathrm{t}$ preserve their equivalent line widths\footnote{The scaling relations are equivalent to the transformed radius measure from \citet{Schmutz1989, Hamann1998}.}. The transformed mass-loss rate is particularly useful for fitting recombination lines in emission. If UV P-Cygni profiles are available, an independent estimate of $\varv_\infty$ can be obtained from the blueward edge of the absorption profile. 

From Eq.\,\eqref{eq: transformed_mdot},  we see that clumping not only affects the recombination line strengths through the explicit $D^{1/2}$ term in the transformed mass-loss rate, but can also change the predicted mass loss and terminal velocity due to ionisation changes in the wind, thereby indirectly affecting the wind strengths. Therefore, in principle, with reliable mass estimates, fitting a hydrodynamic model to observed spectra can provide tight constraints on the clumping stratification (of course, within the stratifications tested).

With the insights gained from the didactic overview, we are now better equipped to understand the full picture of the complexities involved in employing hydrodynamic atmosphere models in quantitative spectroscopy. We revisit the example of increasing $L_\star$ from the beginning of this section, but note that the example discussed here is not just limited to changes in luminosity, but applies to other input parameters as well. Suppose we have a hydrodynamically consistent model whose synthetic spectra provide a decent fit to the observed normalised spectra, including both recombination and P-Cygni lines. However, suppose there is a small mismatch in the absolute flux in the total SED, which could be resolved by increasing $L$. Yet, for dynamically-consistent models increasing $L$ also means an increase of the mass-loss rate and a likely reduction of the terminal velocity. The transformed mass-loss rate increases, and we immediately lose the fit to both types of wind lines. Additionally, as the winds become denser, the difference between the effective temperature at $\tau_\mathrm{Ross} = 2/3$ point and $T_\star$ increases. The electron temperature stratification also changes, affecting the ionization balance and specific temperature-sensitive lines. A potential solution is to increase $T_\star$ in order to restore the $\tau_\mathrm{Ross} = 2/3$ temperature. Increasing $T_\star$ could potentially decrease the mass-loss rate (due to the radius effect; see $\mathrm{Sect.}\,\ref{sec: Hydro_game}$), thereby reducing the difference between the inner boundary and $\tau_\mathrm{Ross} = 2/3$ temperatures. However, the non-monotonic behaviour of mass loss with $T_\star$ means that the mass-loss rate could also increase, further aggravating the issue.

Given these complexities, we did not perform a full quantitative spectroscopic analysis to re-derive stellar parameters for the objects considered in this work as that is not the aim of this work. Stellar parameters such as log$(L_\star/L_\odot)$ and $T_\mathrm{eff}(\tau_\mathrm{Ross} = 2/3)$ can be reliably obtained through non-hydrodynamic models underlying the spectral analysis. The novelty in our work is the introduction of hydrodynamics in the wind, enabling us to predict mass-loss rates by self-consistently modelling the atmospheres. Therefore, we adopt log$(L_\star/L_\odot)$ and $T_\mathrm{eff}(\tau_\mathrm{Ross} = 2/3)$ values from the literature and ensure consistency with the SED by applying similar reddening. 

We begin with the R144 binary system, for which dynamical mass estimates are available. The primary and secondary masses are fixed to {$74$ and $70\,M_\odot$, within error bars of their dynamical estimates}. The H, C, N and O mass fractions are constrained by previous literature estimates. Keeping these inputs fixed, the inner boundary temperature and the clumping stratification are continuously varied until our model $T_\mathrm{eff}(\tau_\mathrm{Ross} = 2/3)$ matches the previously estimated value from the literature, while simultaneously recombination lines of H$\alpha$ and \ion{He}{ii} $\lambda$4686, and UV P-Cygni lines of \ion{C}{iv} $\lambda$1550 are satisfactorily fit where the goodness of fit is judged by eye.

The clumping stratification required for spectral fitting of the R144 binary is then applied to R136a1, where the mass is a priori not known. We fix the clumping stratification and vary the mass until the recombination lines and P-Cygni lines are satisfactorily fit. This is the general strategy; however, we also perform additional testing for R136a1. For example, we fix the mass to the value derived from homogeneous mass relations and adjust the clumping stratification, and also perform tests with varying $\varv_\mathrm{turb}$. Below, we present our spectral fits for the R144 binary system and R136a1, using the technique detailed here.

\section{$\texttt{PoWR}^\textsc{hd}$ models of R144 - primary and secondary}
\label{sec: Hydro_model_spectra_fit_R144}

\begin{figure*}
    \includegraphics[width = \textwidth]{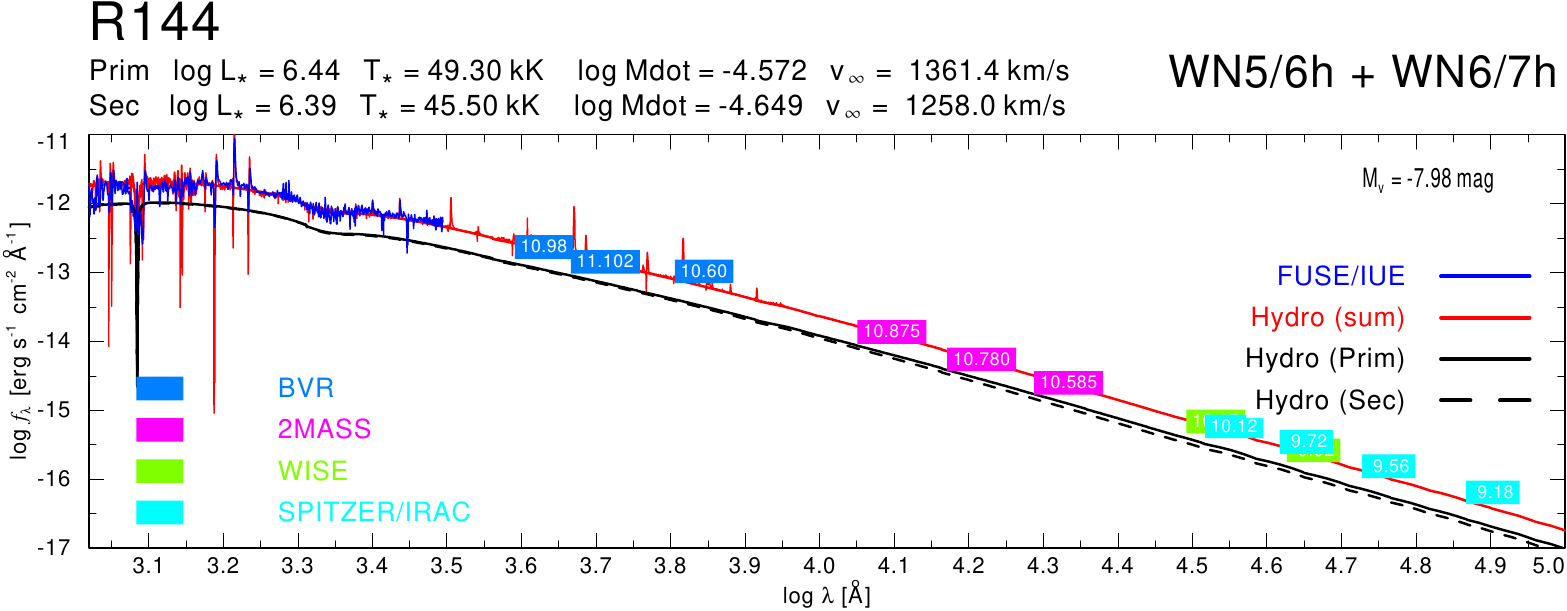}
    \caption{Spectral energy distribution of the R144 binary star system with FUSE/IUE data (blue line), and BVR photometry in the optical and 2MASS, WISE, and SPITZER/IRAC photometry in the infra-red. The red line represents the reddened synthetic spectral energy distribution from our best-fit model. The black solid and dashed lines correspond to the individual contributions from the primary and secondary stars, respectively. }
    \label{fig: R144_sed}
\end{figure*}

\begin{figure}
    \includegraphics[width = \columnwidth]{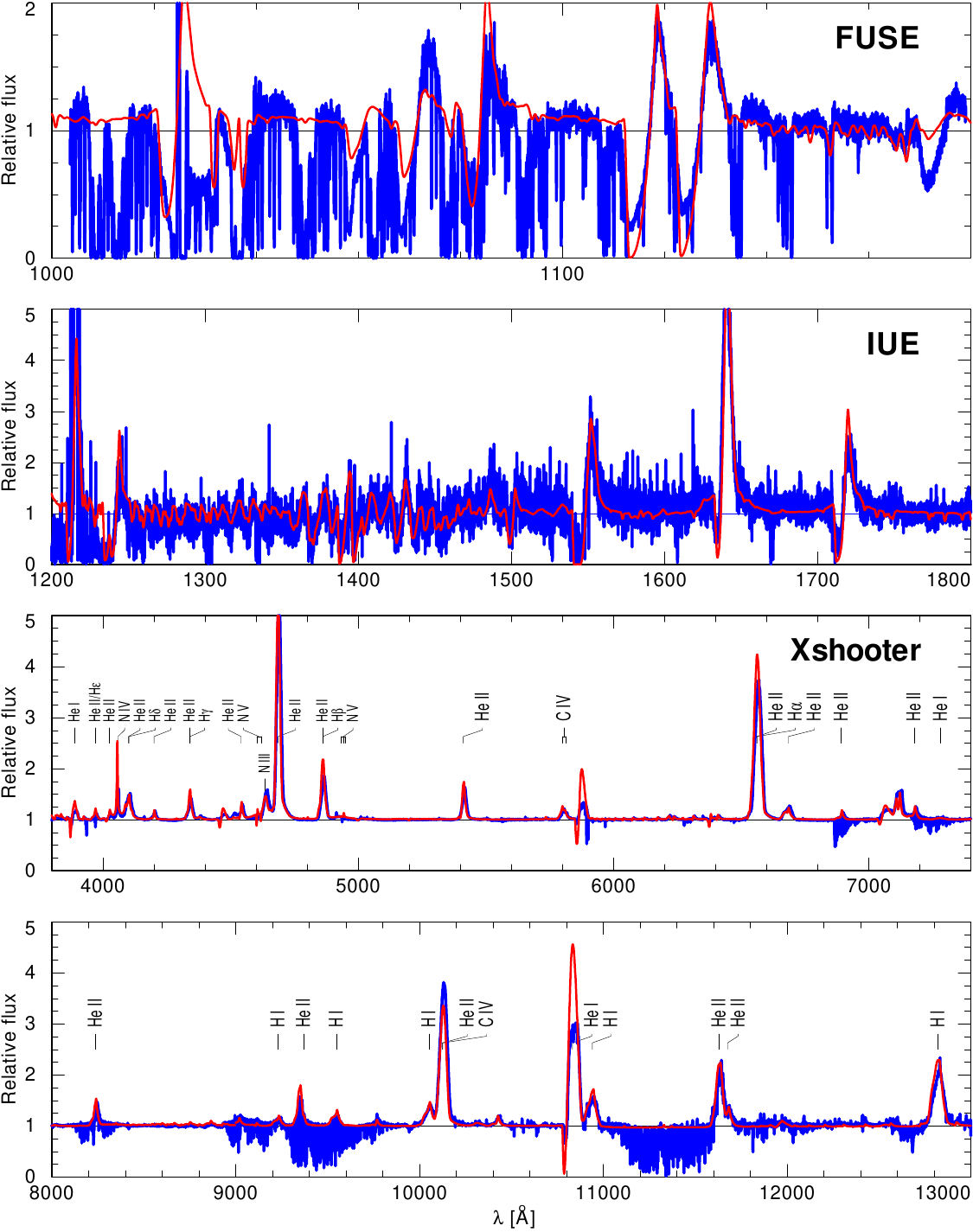}
    \caption{Normalised FUSE, IUE and X-shooter spectra (blue line) compared to the synthetic composite spectra from our best-fit model (red line). The UV FUSE and IUE data are normalised by dividing the absolute flux by the sum of the continuum of the primary and the secondary model. The ground-based X-shooter spectra have not been corrected for H$_2$O and O$_2$ molecular absorption bands in the near IR. }
    \label{fig: R144_norm_tot}
\end{figure}

\begin{figure}
    \includegraphics[width = \columnwidth]{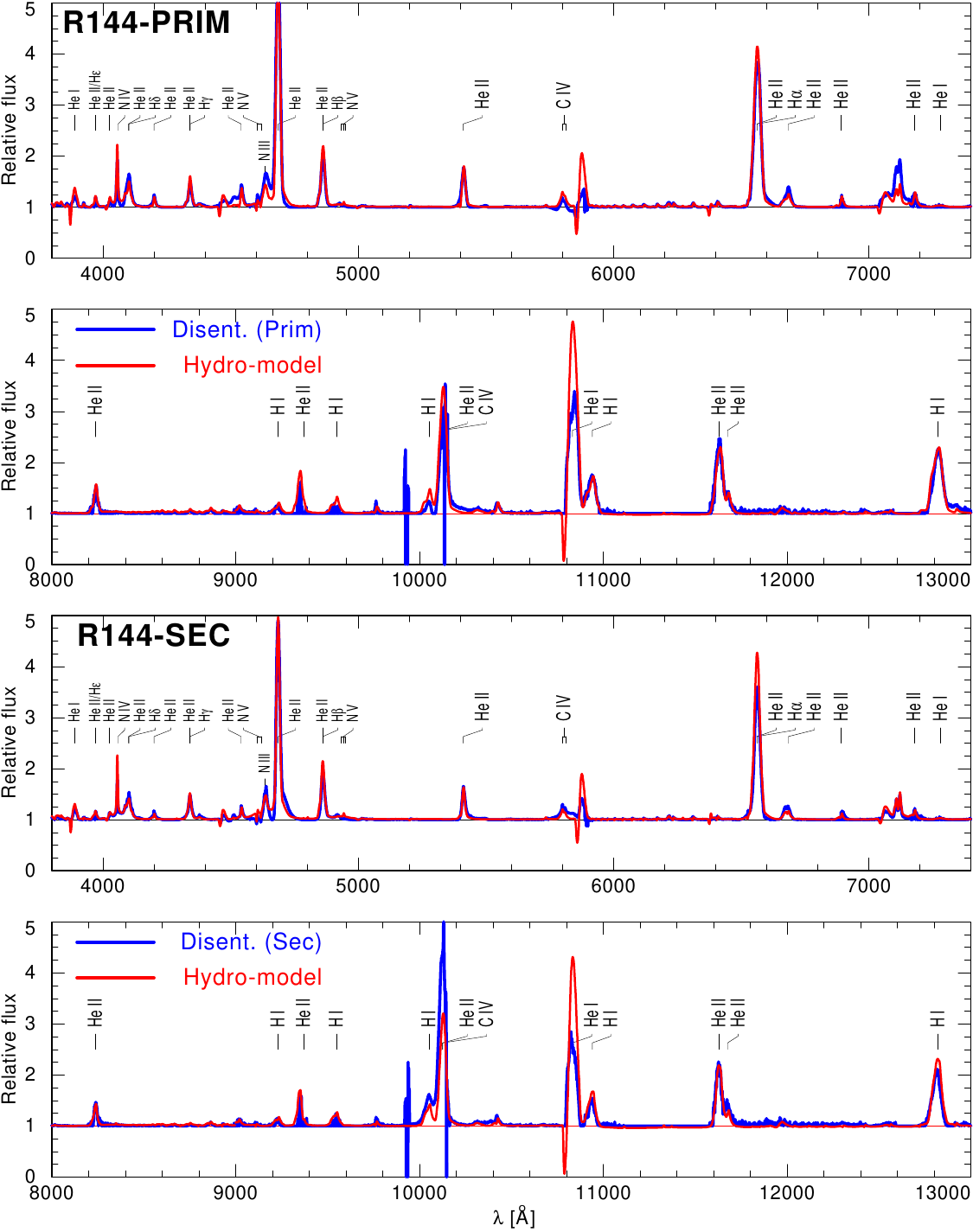}
    \caption{Disentangled spectra of R144 for the hotter primary star and the cooler secondary star (blue lines). The normalised synthetic spectra from our individual hydrodynamic models are shown in red.}
    \label{fig: R144_disent}
\end{figure}

\begin{figure*}
    \includegraphics[width = \textwidth]{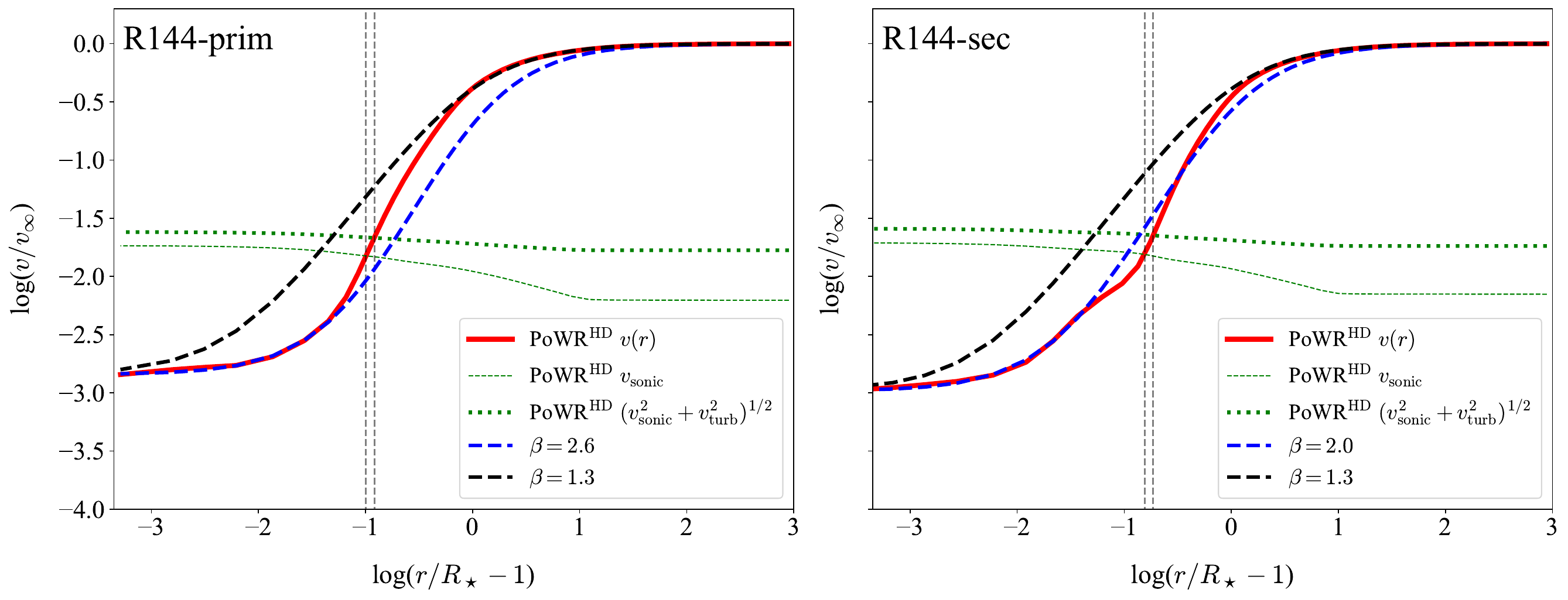}
    \caption{Velocity field (red solid curve) predicted by our best-fit R144 primary and secondary hydrodynamic models. The green dashed and dotted curves represent the isothermal sound velocity and the root-mean-square of the sound and turbulent velocities, respectively. The two vertical black dashed lines indicate the sonic and critical points. Additionally, the black and blue dashed lines correspond to $\beta$-law velocity profiles with values of $\beta$ mentioned in the legend. }
    \label{fig: R144_velo}
\end{figure*}

For the primary and secondary components of R144, we adopted luminosities of $\log(L_\star/L_\odot) = 6.44$ and $6.39$, respectively, as estimated by \citet{Tomer2021}. Dynamical mass estimates of $74\,M_\odot$ and $69\,M_\odot$ for the two components were derived from detailed orbital analyses presented in the same work. \gnstext{The stellar masses in our best-fit R144 models are fixed at $74\,M_\odot$ and $70\,M_\odot$, which are consistent with the dynamical masses within their error bars. These masses are lower compared to estimates from chemical homogeneous assumptions, which along with lower terminal velocities and strong \ion{He}{i} line could hint towards the objects being in their core-He burning phase while still having non-negligible H on top. }

Surface mass fractions of H, C, N, and O were also taken from \citet{Tomer2021}, although the surface H abundance was slightly reduced by 0.1 to achieve a better simultaneous fit to the H/He lines. {The complete abundance distribution for other elements included in modelling are listed in Table \ref{tab: R136_R144_composition} in $\mathrm{Appendix}\,\ref{Appendix: powrHD_model_properties}$, and for them mass fraction spread is set according to solar-scaled abundances from \citet{GS98}.} The turbulent velocity is fixed at $\varv_\mathrm{turb} = 21.21\,\mathrm{km\,s}^{-1}$.

The inner boundary effective temperature, $T_\star$, was adjusted so that the effective temperature at $\tau_\mathrm{Ross} = 2/3$ would closely match the reported values of $T_\mathrm{eff} = 45$ kK and $40$ kK for the primary and secondary components in \citet{Tomer2021}. The resultant fits to the full SED, the normalised composite spectra and the normalised disentangled spectra are shown in $\mathrm{Figs.}\,\ref{fig: R144_sed}$, $\ref{fig: R144_norm_tot}$, and $\ref{fig: R144_disent}$. A distance modulus of 18.48 is adopted \astext{\citep{Pietrzynski2019}}. The reddening applied consists of two components: a Galactic foreground contribution, following \citet{Seaton1979}, with a reddening of $E_{B-V}^\mathrm{MW} = 0.04$ mag and $R_V = 3.1$; and an LMC extinction law derived by \citet{Howarth1983}, with $E_{B-V}^\mathrm{LMC} = 0.15$ mag and $R_V = 3.2$. The total reddening value used is the sum of these contributions: $E_{B-V} = E_{B-V}^\mathrm{MW} + E_{B-V}^\mathrm{LMC} = 0.19$.

\gnstext{Overall, the hydrodynamic model spectra reproduce both the observed SED, as well as the relevant P-Cygni lines in the UV and recombination lines in the optical. One exception to this {are the  \ion{He}{i} $\lambda$5876} and $\lambda$10830 lines, where our models over-predict the emission compared to observations. However, this line can be affected by wind-wind collision, as discussed in \citet{Tomer2021}, and is therefore not considered for our fits. }

The predicted wind properties of our best-fit $\texttt{PoWR}^\textsc{hd}$ models for the primary and secondary components of R144 are presented in the first two columns of $\mathrm{Table}\,\ref{tab: R136_R144_properties}$. The mass-loss rates derived for the two components are $\log(\dot{M}\,[M_\odot/\mathrm{yr}]) = -4.572$ and $-4.649$, respectively. The terminal velocities derived from our models, $1361\,\mathrm{km\,s}^{-1}$ for the primary and $1258\,\mathrm{km\,s}^{-1}$ for the secondary, are in good agreement with the values of $1400\,\mathrm{km\,s}^{-1}$ and $1200\,\mathrm{km\,s}^{-1}$ reported by \citet{Tomer2021}. \gnstext{The clumping factors that we require in our models to simultaneously fit both the recombination and P-Cygni lines are $D_\mathrm{cl,\infty} = 30$ and $22$.} 

Our predicted absolute mass-loss rates are approximately 0.3 dex lower than the empirical rates reported by \citet{Tomer2021}. This difference can be explained by the lower clumping factor of $D_\mathrm{cl,\infty} = 10$ used in their work. For a one-to-one comparison, we can look at the unclumped mass-loss rates (i.e. multiplying by the square root of the clumping factor), we obtain $\log(\dot{M}_\mathrm{unclump}\,[M_\odot/\mathrm{yr}]) = -3.83$ and $-3.97$ for the two components. These values are consistent, within error bars, with the unclumped rates of $-3.88$ and $-3.84$ reported by \citet{Tomer2021}.

Our R144 models favour a clumping stratification that increases outwardly from no clumping up to the above $D_\mathrm{cl,\infty}$ values, \astext{with $\varv_\text{cl} \approx 100\,\mathrm{km\,s}^{-1}$ characterising the onset.} 
Since this onset is beyond the critical point, located at roughly $30\,\mathrm{km\,s}^{-1}$, the clumping stratification has minimal impact on our theoretically predicted mass-loss rates. 

The velocity fields predicted by our best-fit hydro-models are no longer described by a simple $\beta$-law. The actual velocity fields for the two components, normalised to their terminal velocities, are shown in $\mathrm{Fig.}\,\ref{fig: R144_velo}$. We also show the run of the isothermal sound speed and the root-mean-square of the sound and turbulent velocities. The sonic and critical points are identified as the locations where the actual velocity field intersects these curves.

We further overlay $\beta$-law velocity profiles of the form
\begin{equation} \varv(r) \approx \varv_\infty\bigg(1 - f\dfrac{R_\star}{r} \bigg)^\beta,
\label{eq: beta_law}
\end{equation}
where $f$ is adjusted to ensure that the velocity at the inner boundary, $\varv(R_\star)$, matches that of the hydrodynamic model. This particular form of the $\beta$-law provides a smooth transition between the wind domain and the quasi-hydrostatic domain \citep[see][for further details]{Sander2015, Sander2017}. 

In the inner quasi-hydrostatic region, the effective $\beta$ that best captures the velocity profile of our primary (secondary) model is $2.6$ ($2.3$), while in the outer wind region, the effective $\beta$ decreases to $1.3$ ($1.3$). Between these two domains, the velocity field shows a steep increase that cannot be adequately described by a simple $\beta$-parametrisation for the given $\varv_\infty$ and $\varv(R_\star)$. Values of $\beta$ exceeding unity are not uncommon in dynamically consistent VMS models \citep[e.g.][]{Vink2011}.

Regarding the optical thickness of the wind; since our models calculate the flux-weighted mean opacity, we can define the flux-weighted mean optical depth at the sonic point by integrating $\varkappa_{F}$ from the outer boundary inward to the critical point. This provides a measure of the wind's optical thickness. For both the primary and secondary model, we find $\tau_{F,\mathrm{crit}}$ above unity, indicating that the critical point (i.e. the location where the wind is effectively launched) is located in the optically thick regime.

\newgnstext{Finally, we take a look at the H- and He-ionising fluxes. In $\mathrm{Table}\,\ref{tab: R136_R144_properties}$, we give the ionising fluxes of both the R144 components expressed as the logarithm of the number of photons per second blueward of the Lyman continuum edge, the \ion{He}{i} and  \ion{He}{ii} edges. The \ion{H}{i} ionising flux is of the order of $10^{50}-10^{51}$ photons per second, which is an order of magnitude higher than for typical O stars \citep{Smith02} and classical WR stars \citep{Sander2020b}. Only towards the high-mass end ($M_\star > 100\,M_\odot$) do these classical WR stars predict a comparable order of magnitude for \ion{H}{i} ionising flux. However, such high-mass classical WR stars are unlikely to exist. This might indicate VMSs could be formidable sources for the re-ionisation of the Universe \citep[see also][]{Schaerer2025}.}

\newgnstext{The \ion{He}{ii} ionising flux of the R144 components, on the other hand, is roughly 10 orders of magnitude lower compared to the \ion{H}{i} ionising flux and is practically zero. The photon numbers per second are in line with the strong-lined WR stars \citep[see Fig. 22 in][]{Sander2020b}. This is mainly due to the strong, optically thick winds of WNh stars which are not transparent to \ion{He}{ii} ionising photons. This means VMSs are not considered to be the main sources of nebular \ion{He}{ii}. However, narrow \ion{He}{ii} lines could still be arising from the slow stellar winds from VMSs \citep{Graf2015} and they may need to be considered as a potential source of narrow \ion{He}{ii} emission in star-forming galaxies at low $Z$.}

\begin{figure*}
    \includegraphics[width = \textwidth]{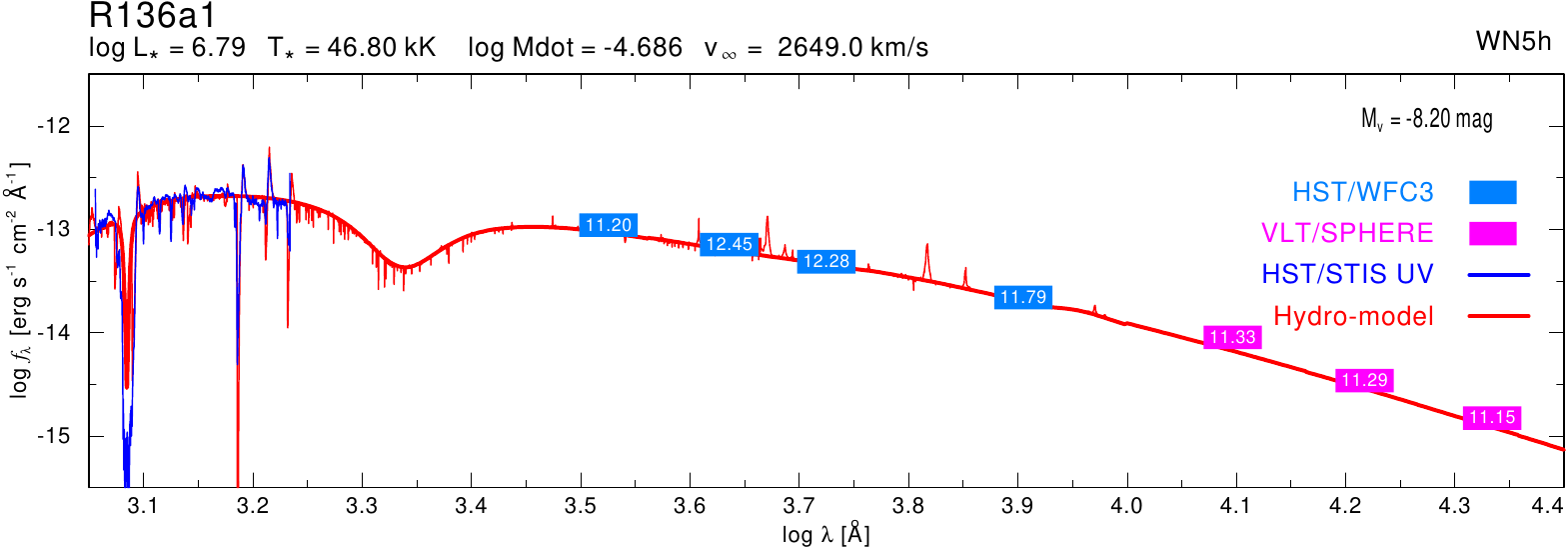}
    \caption{Spectral energy distribution of R136a1 from UV HST/STIS (blue line), and optical and near-IR multi-colour photometry from HST/WFC3 and VLT/SPHERE. The red line is the reddened synthetic spectral energy distribution from our best-fit fixed-clumping R136a1 model. }
    \label{fig: R136a1_sed_clump}
\end{figure*}

\begin{figure}
    \includegraphics[width = \columnwidth]{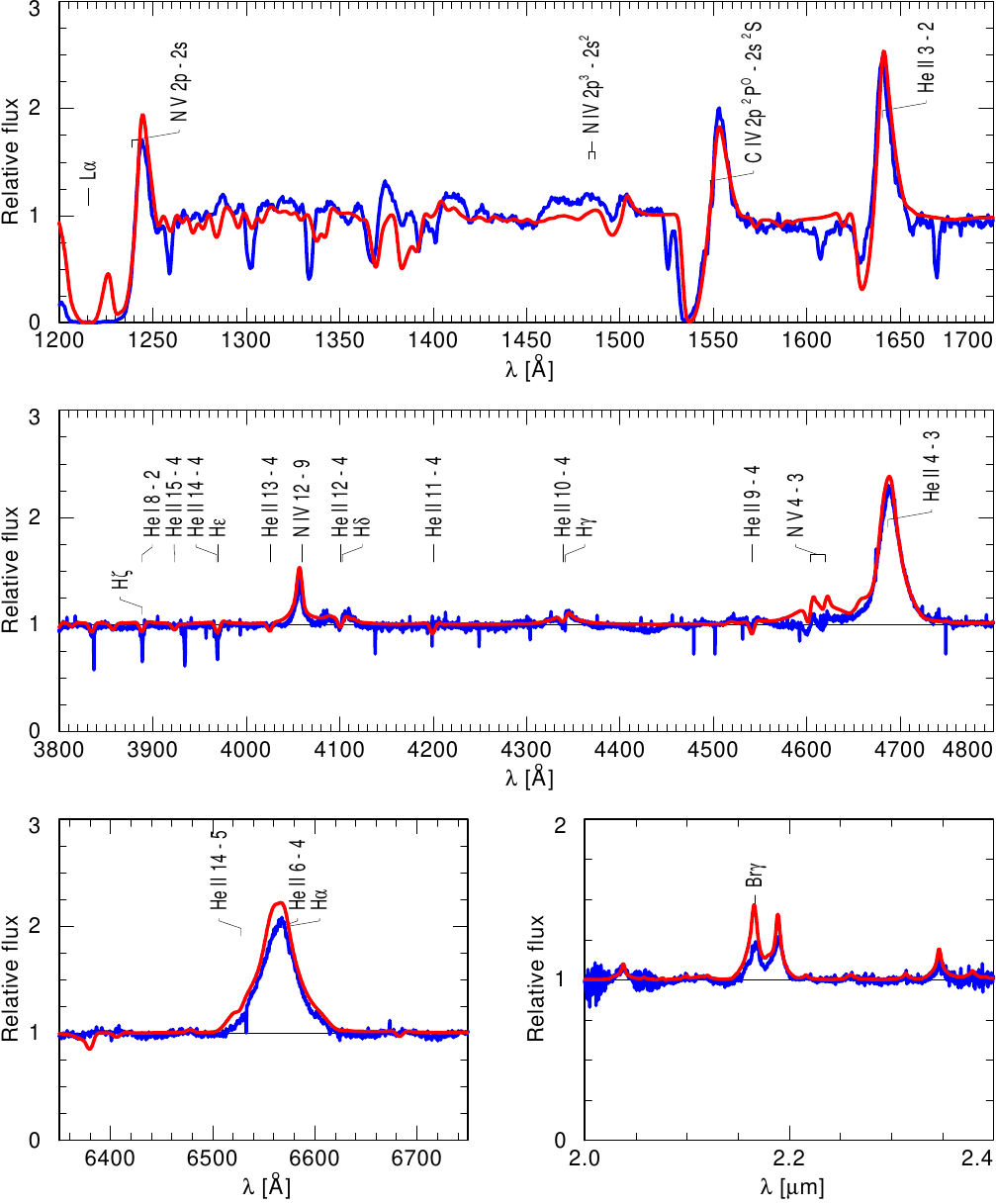}
    \caption{Normalised far-UV HST-STIS/G140L, optical HST-STIS/G430M, H$\alpha$ G750M observations from \citet{Crowther2016}, and K-band VLT/SINFONI observations from \citet{Schnurr2009} in blue. They are compared to the normalised synthetic spectra from our best-fit fixed-clumping R136a1 model (red line). The far-UV HST data are normalised by dividing the absolute flux by the continuum of the model.}
    \label{fig: R136a1_norm_clump}
\end{figure}

\begin{figure}
    \includegraphics[width = \columnwidth]{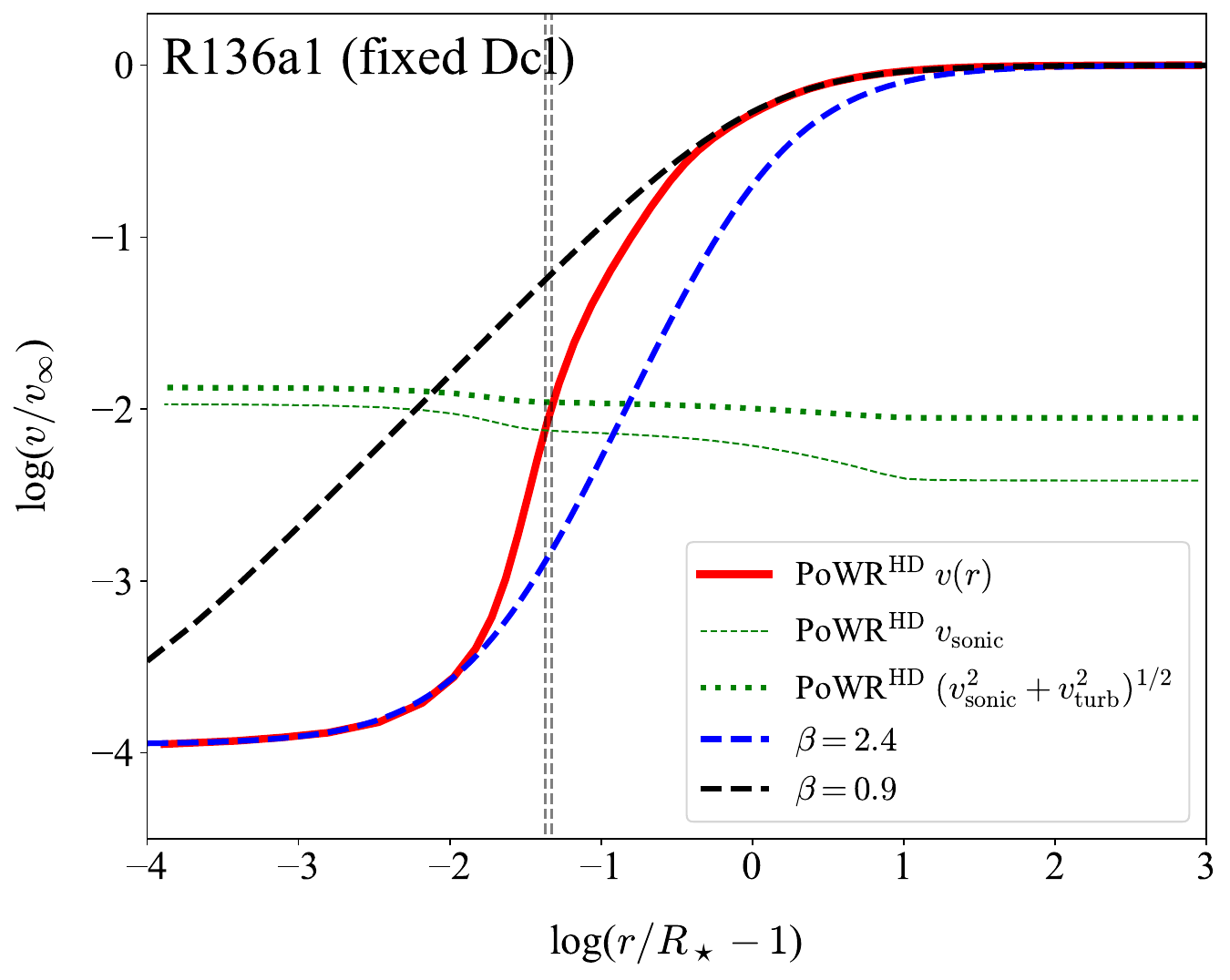}
    \caption{Velocity field (red solid curve) predicted by our best-fit hydrodynamic model for R136a1 (while keeping mass fixed). The black and blue dashed lines correspond to $\beta$-law velocity profiles with $\beta = 0.9$ and $\beta = 2.4$, respectively. All other lines retain the same meaning as $\mathrm{Fig.}\,\ref{fig: R144_velo}$. }
    \label{fig: R136a1_velo}
\end{figure}

\begin{figure*}
    \includegraphics[width = \textwidth]{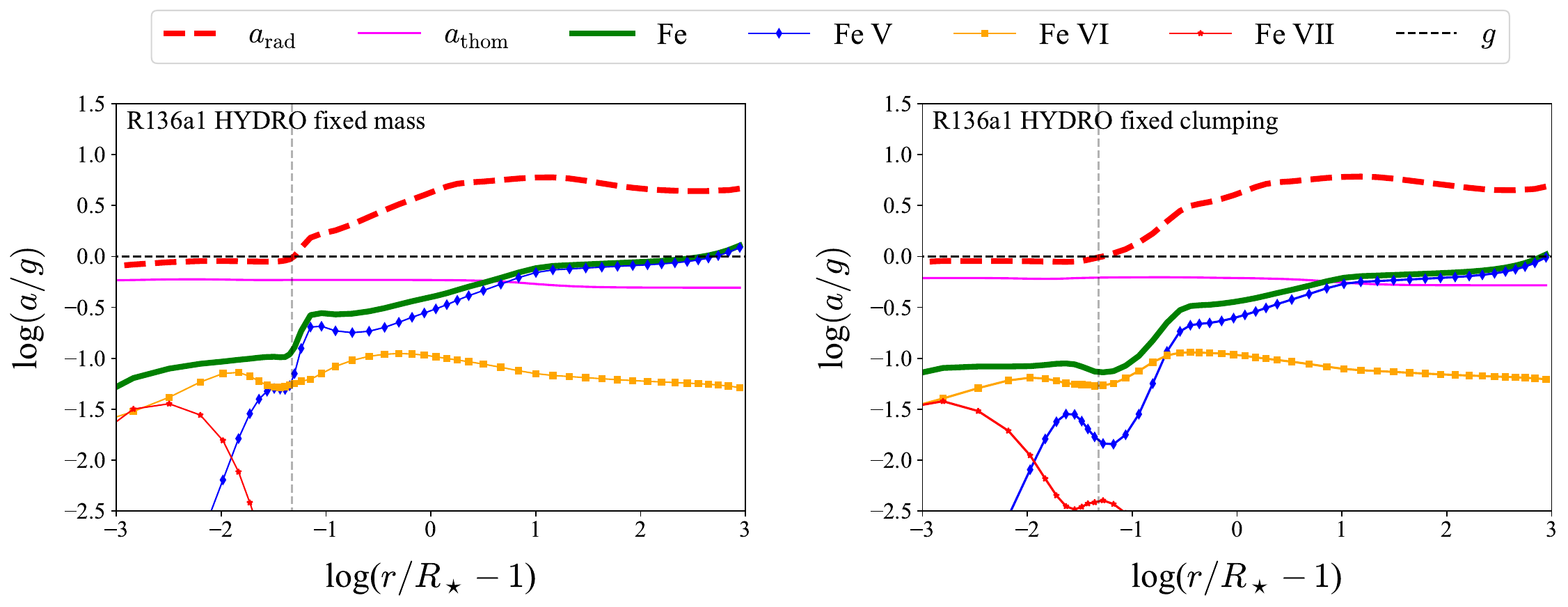}
    \caption{Radiative acceleration stratification of the best-fit hydrodynamic R136a1 models where the mass is fixed to the homogeneous mass (left sub-panel) and where the clumping stratification is fixed to occur outside the critical point (right sub-panel). The individual contributions from lead ions of iron (Fe V, Fe VI, and Fe VII) are shown, along with the total radiative acceleration from iron (green solid line). The total radiative acceleration from all lines and continuum processes is represented by the red dashed line, while the contribution from electron scattering alone is shown in magenta. All accelerations are normalised to gravity (black dashed horizontal line). The critical points of the two models are indicated by gray dashed vertical lines.}
    \label{fig: R136a1_acc}
\end{figure*}

\section{$\texttt{PoWR}^\textsc{hd}$ model of R136a1}
\label{sec: Hydro_model_spectra_fit_R136}

\subsection{Mass prediction for R136a1}
\label{sec: mass_prediction}

In the previous section, we determined the clumping stratification required in our hydrodynamic models by using reliable dynamical mass estimates for R144 and spectral fitting to relevant wind lines. Here, we reverse the approach for R136a1. Specifically, we use the clumping stratification determined above and now treat the inner boundary temperature and \textit{stellar mass} as the varying input parameters, adjusting them until we get a satisfactory fit to the spectra. By coupling atmosphere models that solve the hydrodynamics with simultaneous spectral fitting, we can provide a mass estimate for R136a1.

For R136a1, the luminosity and surface H abundance estimates, $\log(L_\star/L_\odot) = 6.79$ and $X = 0.5$, are adopted from \citet{Best2020}. The total surface metal mass fraction is set to $Z = 0.008$. In VMSs, due to the large convective core {\citep{Yusof2013, Kohler2015, Sabhahit2022},} H-burning products such as He and N can readily reach the surface. For simplicity, we assume the surface abundances closely follow the core and adopt CNO-equilibrium abundances estimated from VMS evolution models {from \citet{Sabhahit2022} using the MESA code \citep{MESA11, MESA13, MESA15, MESA17, MESA19}}, yielding a composition with increased N at the expense of carbon (C) and oxygen (O). The full abundance spread of other metals is detailed in $\mathrm{Appendix}\,\ref{Appendix: powrHD_model_properties}$. The He mass fraction is then determined as $Y = 1 - X - Z$. The turbulent velocity is fixed at $\varv_\mathrm{turb} = 21.21\,\mathrm{km\,s}^{-1}$.

The inner boundary effective temperature and the stellar mass are varied until the effective temperature at $\tau_\mathrm{Ross} = 2/3$ of our model roughly matches the value of $T_\mathrm{eff}(\tau_\mathrm{Ross} = 2/3) = 46$ kK reported in \citet{Best2020}, and a simultaneous fit is obtained for recombination and P-Cygni wind lines. The resulting fits to the full SED and the normalised spectra are shown in $\mathrm{Figs.}\,\ref{fig: R136a1_sed_clump}$ and $\ref{fig: R136a1_norm_clump}$. \astext{Again, a} distance modulus of 18.48 is adopted \astext{\citep{Pietrzynski2019}}. The reddening applied consists of three components: a Galactic foreground contribution, following \citet{Seaton1979}, with a reddening of $E_{B-V}^\mathrm{MW} = 0.04$ mag and $R_V = 3.1$; an LMC extinction law derived by \citet{Howarth1983}, with $E_{B-V}^\mathrm{LMC} = 0.1$ mag and $R_V = 3.2$; and a separate reddening component for R136, with $E_{B-V}^\mathrm{R136} = 0.39$ mag and $R_V = 4$, using the same Seaton law. Higher values of $R_V \approx 4-4.2$ have previously been obtained for the R136 region \citep[see, e.g.][]{Doran2013, Best2020}. The total reddening value used is the sum of these contributions: $E_{B-V} = E_{B-V}^\mathrm{MW} + E_{B-V}^\mathrm{LMC} + E_{B-V}^\mathrm{R136} = 0.53$.

\gnstext{Once again, the spectral fits to the overall SED and relevant wind lines are decent. There is a small discrepancy in the blueward edges of the \ion{C}{iv} $\lambda$1550 P-Cygni line between our model spectra and the observations. While a better fit can be achieved, for instance, by increasing the mass or enhancing clumping in the outer wind--both of which would raise the terminal velocity and better align the blueward edges. However, our goal here is not to achieve perfect spectral fits, but to gain insights about VMS wind structure with the novel hydrodynamic treatment of the winds. }

The mass we predicted from the $\texttt{PoWR}^\textsc{hd}$ models for R136a1 is $M_\mathrm{Hydro} = 233\,M_\odot$. We can compare the hydro-predicted mass with the mass estimated using the chemically homogeneous mass relations from \citet{Graf2011}. Using the luminosity and surface H mass fraction of $\log(L_\star/L_\odot) = 6.79$ and $X = 0.5$, we derive a chemically homogeneous mass of approximately $M_\mathrm{hom} = 242\,M_\odot$. This represents the maximum possible mass from stellar structure for the given luminosity and surface H, assuming the star is fully homogeneous and that the central H abundance equals the surface H abundance. The hydro-predicted mass is lower but still very close to the chemically homogeneous mass estimate. The assumption of chemical homogeneity is reasonable at such high luminosities, as VMS structure suggests that the convective core would occupy nearly the entire stellar interior \citep{Graf2011,Yusof2013,Sabhahit2022}.

\gnstext{The above method of using hydrodynamical models to estimate mass gives a unique value for the mass. This is because even small modifications to the input parameters immediately disrupt the spectral fits. Which means, for a given set of input stellar parameters (other than mass), one can obtain a unique mass for the object with very small error bars by satisfying the hydrodynamics while also fitting the spectra. 
However, the overall precision on the mass value estimated still depends on the errors in the input stellar parameters. 
Performing a detailed error analysis to get the overall error bar on our hydro-predicted mass by varying the different input parameters individually is not possible due to the computational expense of these models.}

The predicted wind properties from our best-fit hydrodynamic model for R136a1, with the clumping stratification fixed, are presented in the third column of $\mathrm{Table}\,\ref{tab: R136_R144_properties}$. The mass-loss rate we predicted is $\log(\dot{M}[M_\odot/\mathrm{yr}]) = -4.686$. Regarding the wind optical depth, we find $\tau_{F,\mathrm{sonic}} \approx 0.76$ which is of the order of unity. To achieve a higher terminal velocity and better fit the P-Cygni lines, we use a slightly higher clumping factor of $D_\mathrm{cl,\infty} = 45$ (compared to the values obtained for R144). By converting to an unclumped mass-loss rate, we obtain $\log(\dot{M}_\mathrm{unclump}[M\odot/\mathrm{yr}]) = -3.86$, which is in agreement with the unclumped rate of $-3.8$ reported by \citet{Best2020}. The terminal velocity we predicted is 2649 km\,s$^{-1}$, which is comparable to the value of 2600 km$^{-1}$ reported by \citet{Best2020}.

More recently, \citet{Brands22} analysed the optical and UV spectroscopy of O and WNh stars in the R136 cluster, including R136a1, using the FASTWIND code coupled with a genetic algorithm. They predicted a mass-loss rate of $\log(\dot{M}[M_\odot/\mathrm{yr}]) = -4.57$. \citet{Brands22} use a different clumping parametrisation with non-void interclump medium, their clumping is  $D_\mathrm{cl,\infty} = 43$.  A first order scaling, which is no longer fully adequate in such cases, would yield an unclumped rate of $\log(\dot{M}_\mathrm{unclump}[M_\odot/\mathrm{yr}]) = -3.75$, which is within $\sim 0.1$ dex of our estimate. 

\newastext{Considering the ionizing fluxes of R136a1, reported also in Table 1,
we find a very similar behaviour as for both R144 components. Notably,
also the wind of R136a1 is not transparent to \ion{He}{ii}-ionizing
photons, despite its $\sim$$0.5\,$dex lower transformed mass-loss rate.
\citet{Sander2023} identified that for WR winds driven by the hot iron
opacity bump, the atmosphere becomes transparent to \ion{He}{ii}
ionizing flux for $\log(\dot{M}_\text{t}\,[M_\odot\,\mathrm{yr}^{-1}])
\leq -4.6$. For WNh-type winds, which are launched further out than
those of the classical WR stars studied in \citet{Sander2023}, this
limit seems to be even lower.}

\subsection{Tests with chemical homogeneous mass}

We also conducted tests by fixing the stellar mass in our atmosphere models to the homogeneous mass of $242\,M_\odot$, while varying the inner boundary temperature and clumping stratification. The predicted wind properties for this fixed-mass model of R136a1 are presented in the fourth column of $\mathrm{Table}\,\ref{tab: R136_R144_properties}$. The corresponding spectral fits are shown in $\mathrm{Appendix}\,\ref{Appendix: spectra_vturb_clump}$. 

We obtained satisfactory fits again, hinting towards a degeneracy between the stellar mass and clumping stratification. As the stellar mass increases from 233 to $242\,M_\odot$, the predicted mass-loss rate initially decreases, and the spectral lines no longer fit satisfactorily. To recover the fits, the mass-loss rate must be increased, which is achieved by having an earlier onset of clumping. With a clumping onset at $\varv_\mathrm{cl} = 23\,\mathrm{km\,s}^{-1}$, the mass-loss rate increases, and the fits are restored. The final predicted mass-loss rate from our fixed-mass R136a1 model is very similar to the values obtained previously.

The low clumping onset in this model places the onset below the critical point, which occurs at roughly $30\,\mathrm{km\,s}^{-1}$. As a result, the clumping assumptions not only affect the terminal velocity but also influence the mass-loss rate. In $\mathrm{Fig.}\,\ref{fig: R136a1_acc}$ (left sub-panel), we show the radial stratification of the acceleration normalised to gravity for our fixed-mass R136a1 model. The impact of having the onset below the critical point is evident in the acceleration stratification, where a surge in radiative acceleration occurs due to Fe lines. This is accompanied by a clear shift in ionization from \ion{Fe}{VI} to \ion{Fe}{V}, which leads to an increase in the predicted mass-loss rate. \newgnstext{Such an increase in the mass-loss rate has previously been reported in the context of bi-stable winds, but at lower temperatures where the ionization switch occurs from \ion{Fe}{iv} to \ion{Fe}{iii} \citep{Vink99}.}

In comparison, the fixed-clumping stratification model from $\mathrm{Sect.}\,\ref{sec: mass_prediction}$ has a delayed onset, meaning the clumping factor near the critical point is nearly unity. This results in a drop in line acceleration at this location, as seen in the right sub-panel of $\mathrm{Fig.}\,\ref{fig: R136a1_acc}$. While there is still a surge in Fe line acceleration, it occurs beyond the critical point, where the iron ionization changes from \ion{Fe}{vi} to \ion{Fe}{v}. In the absence of this ionization switch near the critical point, the radiative acceleration decreases, leading to a drop in the predicted mass-loss rate, which is compensated by the lower stellar mass of $233\,M_\odot$. While both models---fixed clumping stratification with varying mass and fixed mass with varying clumping stratification---can reproduce the observed spectra, the latter has a mass-loss rate that depends on the details of the clumping law used.

\subsection{Tests with different turbulent velocity}
\label{sec: vturb_tests}

Recently, \citet{Debnath2024} performed time-dependent, 2D simulations of O star atmospheres using the hybrid-opacity approach, and found large 2D-averaged turbulent velocities already in the photosphere of their models, of the order of 30-100 km\,s$^{-1}$. Interestingly, the turbulent velocities increase with the Eddington parameter, with their O2 model having $\overline{\varv_\mathrm{turb}} = 100$ km\,s$^{-1}$. We therefore perform additional tests by varying the turbulent velocity in our fixed-mass R136a1 model. 

The predicted wind properties from our turbulence tests with $\varv_\mathrm{turb} = 70.71$ and $100\,\mathrm{km\,s}^{-1}$ are presented in the last two columns of $\mathrm{Table}\,\ref{tab: R136_R144_properties}$. The corresponding spectral fits are shown in $\mathrm{Appendix}\,\ref{Appendix: spectra_vturb_clump}$. As demonstrated in $\mathrm{Sect.}\,\ref{sec: stellar_param_mass_loss}$, the mass-loss rate predicted by our models increases with $\varv_\mathrm{turb}$. To account for this increase in mass loss, the clumping onset is set further out in the wind, as evidenced by the increase in the $\varv_\mathrm{cl}$ with $\varv_\mathrm{turb}$. This compensates for the higher mass loss, ensuring that the final mass-loss rates, terminal velocities, and $D_\mathrm{cl,\infty}$ remain largely unchanged. 

While the clumping onset shifts outwards, it still remains comparable to the critical point. This is because the critical point also moves outward with $\varv_\mathrm{turb}$, as it is located where the wind velocity crosses $\varv(r) = (\varv_\mathrm{sound}^2 + \varv_\mathrm{turb}^2)^{1/2}$.

\begin{table*}
\centering
\renewcommand{\arraystretch}{1.25} 
\caption{Input parameters,  clumping stratification, and  predicted wind properties of our best-fit R144-primary, R144-secondary, and R136a1 hydro-dynamic models. }
\begin{tabular}{l | c c | c | c c c} 
        \hline
        \hline
         Parameter & R144-prim & R144-sec  & R136a1 Fixed $D_\mathrm{cl}$ & & R136a1 Fixed $M$, $\varv_\mathrm{turb}$ tests &    \\
         Spectral type                                 & WN5/6h    & WN6/7h   & WN5h   & WN5h   & WN5h   & WN5h \\
        \hline
         $T_\star$[kK]                                 & 49.3      & 45.5     & 46.8     & 46.8     & 47.5     & 48      \\
         $T_\mathrm{eff}(\tau_\mathrm{Ross}=2/3)$\,[kK]&45.6       & 41.4     & 46       & 46.2     & 46.1     & 45.7    \\
         log($L_\star/L_\odot$)                        & \bf6.44   & \bf6.39  & \bf6.79  & \bf6.79  & \bf6.79  & \bf6.79    \\
         $R_\star/R_\odot$                             & 22.81     & 25.28    & 37.87    & 37.87    & 36.76    & 36      \\
         $R_{\tau_\mathrm{Ross} = 2/3}/R_\odot$        & 26.62     & 30.41    & 38.78    & 38.82    & 39.01    & 39.74   \\
         $M_\star/M_\odot$                             & \bf74     & \bf70    & 233      & \bf242.1 & \bf242.1 & \bf242.1   \\
         $X$                                           & \bf0.25   & \bf0.3   & \bf0.5   & \bf0.5   & \bf0.5   & \bf0.5     \\
         $\varv_\mathrm{turb}$\,[km\,s$^{-1}$]         & 21.21     & 21.21    & 21.21    & 21.21    & 70.71    & 100     \\
         $D_\mathrm{cl,\infty}$                        & 30        & 22       & 45       & 45       & 40       & 42      \\
         $\varv_\mathrm{cl}$\,[km\,s$^{-1}$]           & 90        & 125      & 100      & 23       & 75       & 82      \\
         \hline
         log($\dot{M}$)                                &  $-4.572$ & $-4.649$ & $-4.686$ & $-4.692$ & $-4.747$ & $-4.74$ \\
         $\varv_\infty$[$\,\mathrm{km\,s}^{-1}$]       & 1361.4    & 1257.9   & 2649.03  & 2722.8   & 2699     & 2672.4  \\
         log($\dot{M}_\mathrm{t}$)                 & -4.297    & -4.370   & -4.875   & -4.893   &  -4.970  & -4.947\\
         $\varv(r=r_\mathrm{sonic})$\,[km\,s$^{-1}$]   & 20.53     & 19.53    & 19.99    & 21.07    & 19.1     & 20.98   \\
         $\varv(r=r_\mathrm{crit})$\,[km\,s$^{-1}$]    & 29.27     & 28.46    & 29.06    & 29.82    & 73.14    & 101.73  \\
         $\beta_\mathrm{eff,in}$                       & 2.6       & 2        & 2.4      & 2.3      & 2.9      & 3.3      \\
         $\beta_\mathrm{eff,out}$                      & 1.3       & 1.3      & 0.9      & 0.9      & 1        & 1        \\
         $\Gamma_\mathrm{e,R_\star}$                   & 0.713     & 0.701    & 0.608    & 0.574    & 0.591    & 0.588   \\
         $\tau_\mathrm{F,sonic}$                       & 1.779     & 1.596    & 0.767    & 0.693    & 0.778    & 0.859   \\
         $\tau_\mathrm{F,crit}$                        & 1.592     & 1.383    & 0.734    & 0.666    & 0.576    & 0.541   \\         
         $\mathrm{log}(Q_{\mathrm{H}\, \textsc{i}}[\,\mathrm{phot\,s}^{-1}])$   & 50.30     & 50.25    & 50.66    & 50.65    & 50.66    & 50.67 \\
         $\mathrm{log}(Q_{\mathrm{He}\, \textsc{i}}[\,\mathrm{phot\,s}^{-1}])$  & 49.73     & 49.62    & 50.10    & 50.07    & 50.12    & 50.14 \\
         $\mathrm{log}(Q_{\mathrm{He}\, \textsc{ii}}[\,\mathrm{phot\,s}^{-1}])$ & 40.17     & 40.05    & 41.17    & 41.23    & 41.20    & 41.24 \\
         
        \hline
\end{tabular}
\tablefoot{The inputs we adopt from previous literature are marked in bold-faced text. For R144-primary and secondary components, we adopt the luminosities and the dynamical masses from \citet{Tomer2021}. The surface H for R144 were also initially adopted from \citet{Tomer2021}, but was later reduced by 0.1 to achieve better fits to H/He lines. For R136a1, the luminosities and surface H 
are adopted from \citet{Best2020}, and the masses are from homogeneous relations from \citet{Graf2011}.}
\label{tab: R136_R144_properties}
\end{table*}

\begin{figure*}
    \includegraphics[width = \textwidth]{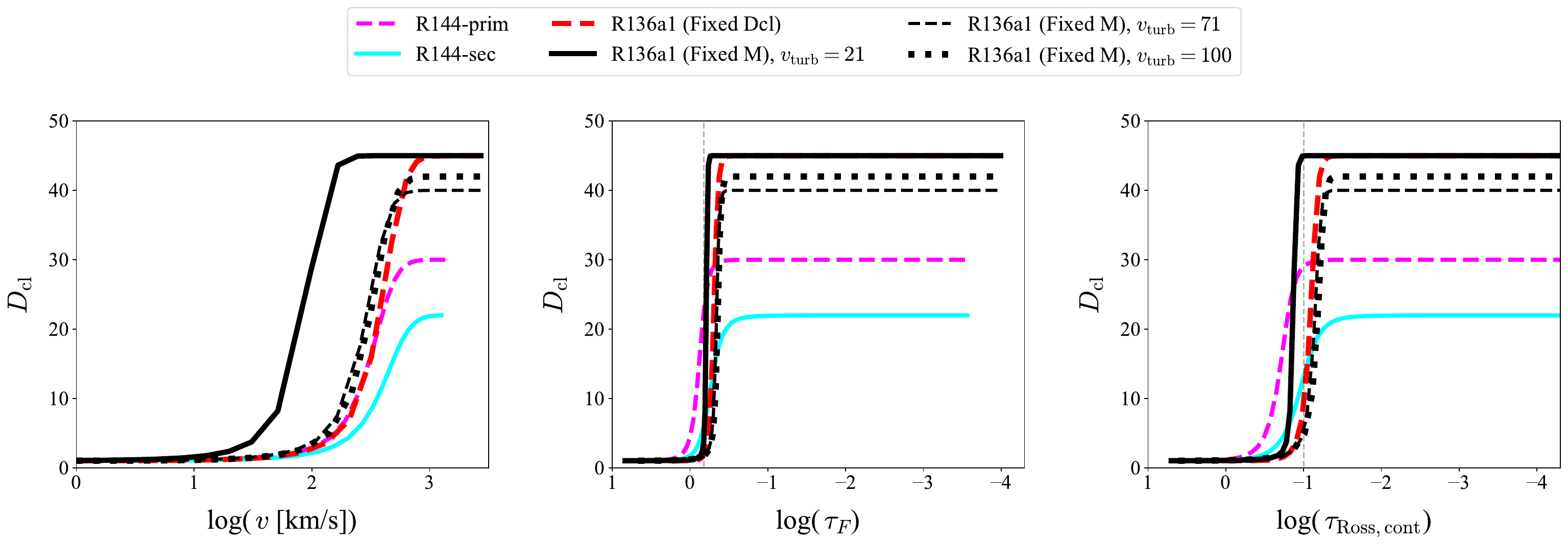}
    \caption{Radial stratification of clumping used in our best-fit hydrodynamic models as a function of velocity (left sub-panel) and flux-weighted mean optical depth (right sub-panel). }
    \label{fig: clumping_strat_all}
\end{figure*}

\section{Discussion}
\label{sec: discussion}

Here, we discuss the advantages and caveats of our work. One of the fundamental challenges in mass-loss diagnostics based solely on recombination lines in emission is the degeneracy between mass-loss rate and clumping. For a given H$\alpha$ line strength, any combination of $\dot{M}$ and $D_\mathrm{cl,\infty}$ that satisfies the relationship $\dot{M} \times D_\mathrm{cl,\infty}^{1/2}$—producing a similar transformed mass-loss rate—will fit the H$\alpha$ line. This assumes that $\varv_\infty$ and $L$ are reasonably well-constrained by the P-Cygni lines and the SED. Consequently, higher clumping in models leads to a lower empirically derived mass-loss rate.

However, a multi-wavelength spectral fit using hydrodynamically consistent models can technically resolve this degeneracy. Any changes to the clumping stratification (or to other input parameters, for that matter) will disrupt the overall spectral fit. For example, alterations in clumping may affect the predicted terminal velocity, causing the P-Cygni blueward edge to no longer align, or the predicted emission line strengths may no longer match the observations.

We observe certain common features in the clumping stratification ultimately used in our models to match the spectra. In all cases where the mass was fixed and clumping was allowed to vary, our models strongly favour a solution in which the clumping starts from unity (no clumping) and then \textit{increases} outward. As shown in $\mathrm{Sect.}\,\ref{sec: clumping_effect}$, if clumping is already significant near the critical point, the mass-loss rate increases while the terminal velocity decreases. This suggests that an even higher clumping factor may be required in the outer wind than what we currently have, meaning the stratification would still increase outward.

A second common feature can be seen in $\mathrm{Fig.}\,\ref{fig: clumping_strat_all}$, where we plot clumping stratification as a function of $\log(\varv)$, the flux-weighted mean optical depth {and the Rosseland continuum mean optical depth} for all models listed in $\mathrm{Table}\,\ref{tab: R136_R144_properties}$. There is a scatter of more than one dex in the location where the clumping onset occurs in velocity space (left panel). However, in the middle and right sub-panels, where we change the abscissa to flux-weighted mean optical depth {and Rosseland continuum mean optical depth respectively}, we observe a consistent feature across all the models tested---ranging from R144 to R136a1 and from varying $\varv_\mathrm{turb}$ to varying clumping stratification. That is, the clumping onset in optical depth space aligns at roughly {a flux-weighted mean optical depth} of $2/3$ {and a Rosseland continuum mean optical depth of 0.1}.

While we can constrain the stratification of clumping from R144 models, we find the relatively high final $D_\mathrm{cl,\infty}$ result  requires a cautious approach. From our R136a1 tests, where clumping was fixed and mass was allowed to vary, we find that a satisfactory fit can be achieved even when clumping remains close to unity in the quasi-hydrostatic region of the model. Therefore, the high clumping values primarily arise from the need to match the terminal velocity using the blueward P-Cygni edge of the objects.

The predicted terminal velocities in our models, however, are highly sensitive to various factors. For example, if our abundances are inaccurate (particularly for elements that dominate radiative driving in the outer wind, e.g. C, N, O, and Ar), the predicted terminal velocity could differ \citep{Vink99}. Another source of uncertainty is the temperature stratification in the outer wind, where our models can predict a non-monotonic stratification, 
\astext{whereas current multi-D models do not show such non-monotonicity (González-Torà et al. 2024) but so far assume radiation and gas temperature to be identical \citep{Moens2022,Debnath2024}.}
As a result, the terminal velocities could show large scatter, due to both uncertain inputs (such as abundances and turbulence) and missing physics in our 1D framework. Therefore, while our models clearly favour a clumping stratification with outward-increasing behaviour with an onset aligned in $\tau$-space over an outward-decreasing stratification, it would be premature to rule out the possibility of an initial outward increase to high values, followed by a subsequent outward-decreasing stratification, \astext{as recently hinted by studies of IR and radio observations of OB stars \citep{Puls2006, Najarro2009, Rubio-Diez2022}.} 

\section{Overview and conclusions}
\label{sec: conclusions}

In this work, we present the first ever hydrodynamical atmosphere models of R144, a WNh+WNh binary star system in the LMC, and R136a1, the most luminous very massive star in the Local Group.  We used the hot-star, nLTE expanding atmosphere code PoWR, with updates to solve stationary wind hydrodynamics. Hydrodynamic atmosphere modelling with simultaneous fits to multi-wavelength spectra in principle can give us the stellar mass, as mass enters the hydrodynamic equation of motion directly affecting the mass loss and wind line strengths. However, a big uncertainty in our models is the clumping stratification. To tackle this, we first make use of reliable dynamical mass estimates available for R144 to first inform us about the clumping stratification required by our models to fit the relevant wind lines. We then reverse the approach for R136a1, that is, we keep use of the clumping stratification obtained above while varying the mass, to estimate a mass for R136a1 using a completely new and independent way.

To perform spectral fitting with hydrodynamic models, we first provide a didactic overview of how different input parameters influence the mass-loss rate, the terminal velocity, and, consequently, the emission line strengths of recombination and P-Cygni lines. 

In general, the variation in mass-loss rates and wind velocities can be understood by analysing the force conditions below and above the critical point, where the flow velocity crosses the isothermal sound speed corrected for turbulent velocity.  If the forces in the inner quasi-hydrostatic, sub-critical region increase relative to gravity, the mass-loss rate increases, and vice versa. Similarly, if the forces in the outer, super-critical wind region increase, the terminal velocity rises, and vice versa.

To connect the wind properties to the line emission measure, we use the so-called transformed mass-loss rate introduced in \citet{Grafener2013}. Models with different combinations of mass-loss rate, terminal velocity, luminosity, and clumping, but with similar transformed mass-loss rates, conserve their line equivalent widths. The transformed mass-loss rate is particularly useful for fitting recombination lines in emission. Additional constraints on the terminal velocity can be obtained using the blueward edge of the UV P-Cygni profiles. 

The general strategy we use for hydrodynamical modelling of R144 components is as follows: we fix the input stellar parameters such as luminosity, effective temperature at $\tau_\mathrm{Ross} = 2/3$, chemical abundances from previous literature estimated with non-hydrodynamic atmosphere models. The masses are fixed to their dynamical mass estimates from \citet{Tomer2021}, while the inner boundary temperature and clumping stratification are continuously varied until the models satisfactorily fit the recombination lines, such as H$\alpha$ and \ion{He}{ii} $\lambda$4686, and align with the blueward edge of the \ion{C}{iv} $\lambda$1550 P-Cygni line. 

\vspace{0.2 cm}
\noindent
We predicted the following wind properties for R144:
\begin{enumerate}
\item The best-fit hydrodynamic models for the primary and secondary components predict mass-loss rates of $\log(\dot{M}\,[M_\odot/\mathrm{yr}]) = -4.572$ and $-4.649$, respectively. 
\item Our models favour an outward-increasing clumping stratification. The clumping goes from no clumping to the maximum values with an onset at roughly $100\,\mathrm{km\,s}^{-1}$ for both components.
\item Relatively high final clumping factors are obtained: 30 for the primary and 22 for the secondary.
\item The velocity stratification is no longer described by a single $\beta$-law. The effective $\beta$ values in the inner quasi-hydrostatic regime are high (2-2.6) and close to unity in the outer wind.
\item The wind optical depths for both components are of the order of unity.
\item The hydrogen-ionising flux of both components is an order of magnitude higher compared to typical O stars and classical WR stars, which might indicate VMSs could be formidable sources for the re-ionisation of the Universe. However, the winds are not transparent to \ion{He}{ii}-ionising flux due to the strong, optically thick nature of the winds.
\end{enumerate}

We now use this clumping stratification that increases outward with an onset of $100\,\mathrm{km\,s}^{-1}$ to model R136a1. The clumping value at the outer boundary is later varied to get a better agreement for the terminal velocity of R136a1 with the blueward edge of the \ion{C}{iv} $\lambda$1550 P-Cygni line. By doing so, we can estimate a mass that is consistent with the wind hydrodynamics in our models while simultaneously fitting the relevant wind lines.
We predicted the following wind properties and stellar mass for R136a1:
\begin{enumerate}
\item The best-fit hydrodynamic model, with fixed-clumping stratification and varying mass, predicts a mass-loss rate of $\log(\dot{M}\,[M_\odot/\mathrm{yr}]) = -4.686$ and needs a relatively high clumping factor of $D_\mathrm{cl,\infty} = 45$.
\item The mass we predicted for R136a1 using our hydrodynamic atmosphere models is $M_\mathrm{Hydro} = 233\,M_\odot$, which is comparable {to the chemical homogeneous mass estimate of $M_\mathrm{hom} = 242\,M_\odot$ using mass-luminosity relations from \citet{Graf2011}}.
\item  Similar to R144, we find a high effective $\beta = 2.4$ in the inner quasi-hydrostatic region, transitioning to an effective $\beta$ close to unity in the outer wind.
\item The wind optical depth is of the order of unity, which is more optically thick compared to typical O-star winds.
\item In fixed-mass R136a1 models, where the mass is fixed using homogeneous mass relations from \citet{Graf2011} and clumping is varied, the clumping onset occurs near the critical point, affecting the \ion{Fe}{vi} to \ion{Fe}{v} ionization change. The clumping details in such models can affect the predicted mass-loss rate.
\end{enumerate}

There are certain common features in the clumping stratification ultimately used in our models to match the spectra. In all cases, where the mass was fixed and clumping was allowed to vary, our models favour a solution where the clumping increases outward. Across all the models presented in this work, ranging from varying $\varv_\mathrm{turb}$ to fixing clumping and varying mass, and from R144 to R136a1, the clumping onset consistently aligns at a flux-weighted mean optical depth of approximately $2/3$, while the onset shows more than one dex scatter in velocity space.

While our models require high clumping in the outer wind, this is primarily to achieve the correct terminal velocities needed to match the P-Cygni profiles. If the observed terminal velocities could be reproduced without invoking high clumping in the outermost part of the wind, either due to changes in certain metal abundances in the models or multi-D effects that are not considered here, then we could technically also fit the spectra with a model with clumping stratification that initially increases outward reaching a maximum value and then decreases outward. Based solely on our models, it would be premature to rule out this possibility.

\begin{acknowledgements}
We thank the anonymous referee for constructive comments that helped improve the paper. This work made use of the Potsdam Wolf-Rayet (PoWR) model atmospheres. The PoWR code, as well as the associated post-processing and visualisation
tool WRplot, has been developed under the guidance of Wolf-Rainer Hamann with substantial contributions from Lars Koesterke, Götz Gräfener,
Andreas Sander, Tomer Shenar and other co-workers and students. GNS and JSV are supported by STFC funding under grant number ST/Y001338/1. AACS, MBP, and RRL are supported by the Deutsche Forschungsgemeinschaft (DFG - German Research Foundation) in the form of an Emmy Noether Research Group – Project-ID 445674056 (SA4064/1-1, PI Sander). TS acknowledges support by the Israel Science Foundation (ISF) under grant number 0603225041
\end{acknowledgements}

%
%

\bibliographystyle{aa}
\bibliography{References.bib}

\appendix

\onecolumn
\section{Wind properties and abundances of our $\texttt{PoWR}^\textsc{hd}$ models}
\label{Appendix: powrHD_model_properties}

\begin{figure}[h]
    \includegraphics[width = \textwidth]{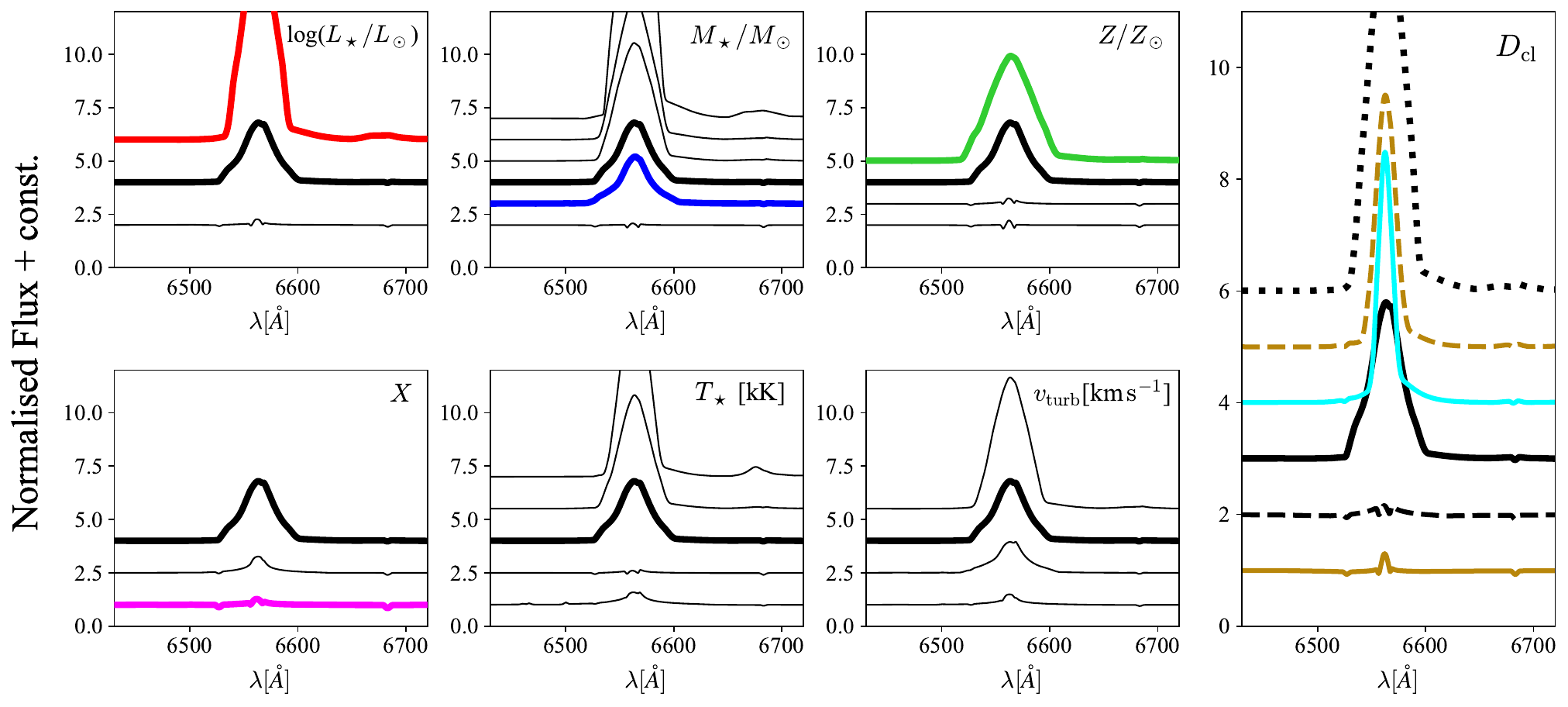}
    \caption{Synthetic H$\alpha$ recombination line from hydrodynamic atmosphere models. The black solid line represents the `base model' from Sect. \ref{sec: Hydro_game}. Each sub-plot illustrates the effect of individually varying input parameters (shown in top left corner) on the strength of the synthetic H$\alpha$ line. The colour scheme corresponds to $\mathrm{Figs.}\,\ref{fig: hydro_game}$ and $\ref{fig: denscon}$.}
    \label{fig: hydro_game_spectra_Halpha}
\end{figure}

\begin{figure}[h]
    \includegraphics[width = \textwidth]{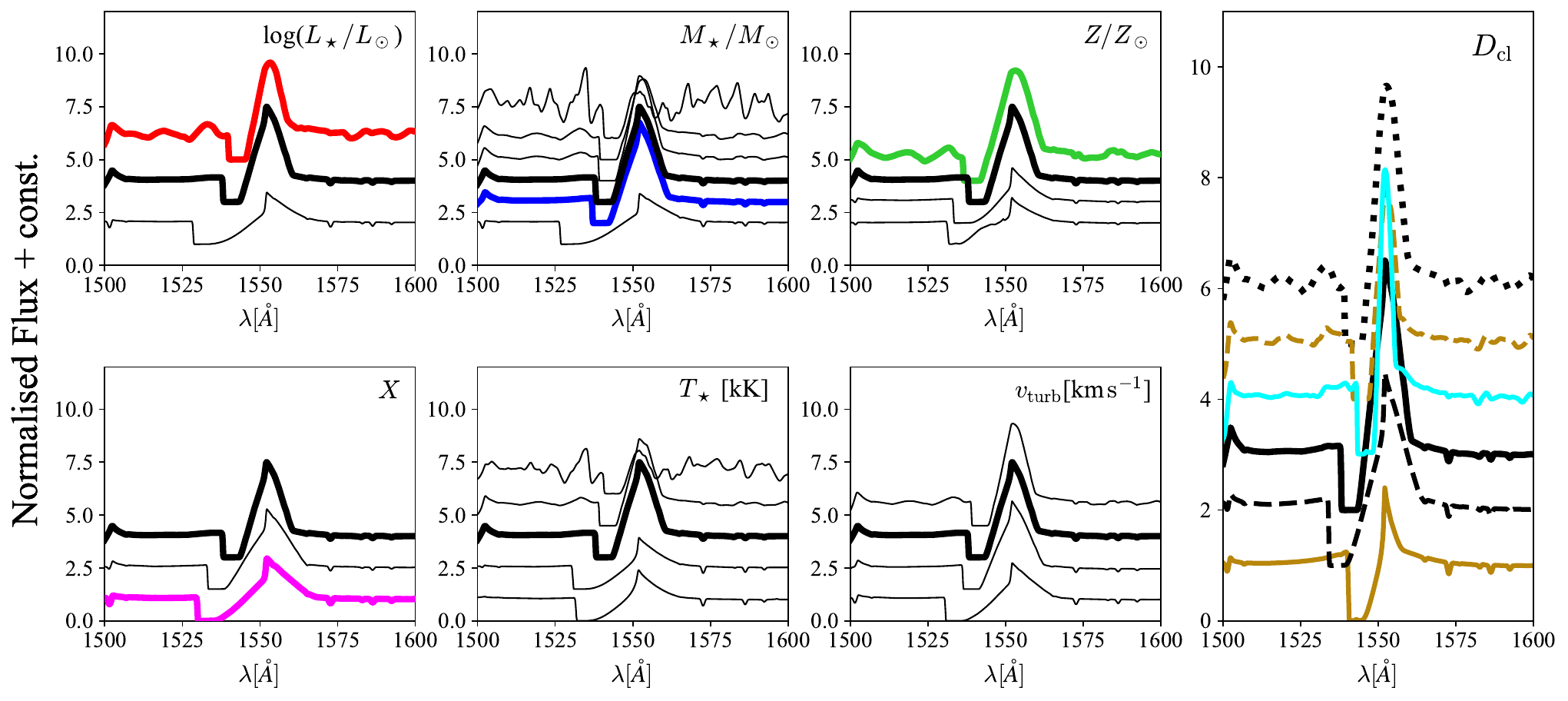}
    \caption{Synthetic \ion{C}{iv} $\lambda$1550 P-Cygni line from hydrodynamic atmosphere models. The black solid line represents the `base model' from Sect. \ref{sec: Hydro_game}. Each sub-plot illustrates the effect of individually varying input parameters (shown in top left corner) on the width of the synthetic \ion{C}{iv} $\lambda$1550 line. The colour scheme corresponds to $\mathrm{Figs.}\,\ref{fig: hydro_game}$ and $\ref{fig: denscon}$.}
    \label{fig: hydro_game_spectra_UV}
\end{figure}

In $\mathrm{Sect.}\,\ref{sec: stellar_param_mass_loss}$, we conducted a didactic study on the effect of different parameters on the wind properties by individually changing parameters from a `base model'. In $\mathrm{Table}\,\ref{tab: wind_prop_PoWR_HD}$, we provide a list of input stellar parameters and the predicted wind properties. {In $\mathrm{Figs.}\,\ref{fig: hydro_game_spectra_Halpha}$ and $\ref{fig: hydro_game_spectra_UV}$, we show the normalised H$\alpha$ and \ion{C}{iv} $\lambda$1550 P-Cygni line spectrum obtained for these models.}

In $\mathrm{Table}\,\ref{tab: base_model}$ we also list the mass fractions of the `base model' and the model sequence where the total $Z$ was modified. The abundance distribution of individual metals follows the solar-scaled composition from \citet{GS98}.

In $\mathrm{Table}\,\ref{tab: R136_R144_composition}$, we present the mass fractions for all elements used in our $\texttt{PoWR}^\textsc{hd}$ models for $\mathrm{R136a1}$ and the $\mathrm{R144}$ binary. For $\mathrm{R136a1}$, the hydrogen and total metal mass fractions are fixed at $X = 0.5$ and $Z = 0.008$. The abundance distribution of individual metals, except for the CNO elements, follows the solar-scaled composition from \citet{GS98}. For the CNO abundances, we adopt N enhancement at the expense of C and O.

\begin{sidewaystable*}
\centering
\renewcommand{\arraystretch}{1.3} 
\begin{tabular}{c c c c c c c c c c c c c c c c c c c} 
        \hline
        & log($L_\star/L_\odot$) & $T_\star$[kK] & $M_\star/M_\odot$ & $X$ & $Z$ & $\varv_\mathrm{turb}$[$\mathrm{km\,s}^{-1}$] & $D_\mathrm{cl,R_\star}$ & $D_\mathrm{cl,\infty}$ & $\varv_\mathrm{cl}$[$\mathrm{km\,s}^{-1}$] & $\varv_\mathrm{cl,end}$[$\mathrm{km\,s}^{-1}$] & log($\dot{M}$) & $\varv_\infty$[$\mathrm{km\,s}^{-1}$] & log($\dot{M}_t$) \\
        \hline
        Base model & 6.4 & 50 & 100 & 0.736 & 0.008 & 70.71 & 1.02 & 25 & 100 & - & $-4.866$ & 1720.22 & $-4.702$           \\
        $L_\star$ sequence & 6.35 & 50 & 100 & 0.736 & 0.008 & 70.71 & 1.0 & 25 & 100 & - & $-5.557$ & 3659.67 & $-5.684$   \\ 
        & 6.45 & 50 & 100 & 0.736 & 0.008 & 70.71 & 1.23 & 25 & 100 & - & $-4.312$ & 1347.47 & $-4.080$                     \\
        $T_{\star}$ sequence & 6.4 & 60 & 100 & 0.736 & 0.008 & 70.71 & 1.01 & 25 & 100 & - & $-5.316$ & 3073.28 & $-5.405$ \\
        & 6.4 & 55 & 100 & 0.736 & 0.008 & 70.71 & 1.01 & 25 & 100 & - & $-5.405$ & 3260.86 & $-5.519$                      \\
        & 6.4 & 45 & 100 & 0.736 & 0.008 & 70.71 & 1.04 & 25 & 100 & - & $-4.604$ & 1491.62 & $-4.379$                      \\ 
        & 6.4 & 40 & 100 & 0.736 & 0.008 & 70.71 & 1.08 & 25 & 100 & - & $-4.307$ & 1236.94 & $-4.001$                        \\
        $M_\star$ sequence & 6.4 & 50 & 85 & 0.736 & 0.008 & 70.71 & 1.21 & 25 & 100 & - & $-3.968$ & 1274.19 & $-3.675$    \\ 
        & 6.4 & 50 & 90 & 0.736 & 0.008 & 70.71 & 1.11 & 25 & 100 & - & $-4.409$ & 1452.81 & $ -4.173$                       \\
        & 6.4 & 50 & 95 & 0.736 & 0.008 & 70.71 & 1.07 & 25 & 100 & - & $-4.523$ & 1516.28 & $-4.305$                       \\
        & 6.4 & 50 & 105 & 0.736 & 0.008 & 70.71 & 1.01 & 25 & 100 & - & $-4.98$ & 1931.92 & $-4.867$                       \\
        & 6.4 & 50 & 110 & 0.736 & 0.008 & 70.71 & 1.0 & 25 & 100 & - & $-5.557$ & 4033.63 & $-5.764$                       \\
        $X$ sequence & 6.4 & 50 & 100 & 0.6 & 0.008 & 70.71 & 1.01 & 25 & 100 & - & $-5.238$ & 2703.51 & $ -5.271$           \\
        & 6.4 & 50 & 100 & 0.5 & 0.008 & 70.71 & 1.0 & 25 & 100 & - & $-5.41$ & 3368.71 & $-5.539$                          \\
        $Z$ sequence & 6.4 & 50 & 100 & 0.736 & 0.02 & 70.71 & 1.06 & 25 & 100 & - & $-4.451$ & 2021.24 & $ -4.358$          \\
        & 6.4 & 50 & 100 & 0.736 & 0.004 & 70.71 & 1.0 & 25 & 100 & - & $-5.454$ & 2714.54 & $-5.488$                       \\
        & 6.4 & 50 & 100 & 0.736 & 0.002 & 70.71 & 1.0 & 25 & 100 & - & $-5.783$ & 3163.28 & $-5.885$                       \\
        $\varv_\mathrm{turb}$ sequence & 6.4 & 50 & 100 & 0.736 & 0.008 & 100 & 1.07 & 25 & 100 & - & $-4.511$ & 1523.79 & $-4.295$    \\ 
        & 6.4 & 50 & 100 & 0.736 & 0.008 & 49.5 & 1.01 & 25 & 100 & - & $-5.035$ & 2075.03 & $-4.953$                       \\
        & 6.4 & 50 & 100 & 0.736 & 0.008 & 35.36 & 1.0 & 25 & 100 & - & $-5.36$ & 3203.94 & $-5.467$                        \\  
        $D_\mathrm{cl}$ sequence & 6.4 & 50 & 100 & 0.736 & 0.008 & 70.71 & 1.0& 1 & - & - & $-5.289$ & 1367.98 & $ -5.725$  \\
        & 6.4 & 50 & 100 & 0.736 & 0.008 & 70.71 & 4.0 & 4 & - & - & $-4.532$ & 993.72 & $-4.528$                           \\
        & 6.4 & 50 & 100 & 0.736 & 0.008 & 70.71 & 1.0 & 25 & 300 & - & $-5.256$ & 2527.29 & $-5.261$                       \\
        & 6.4 & 50 & 100 & 0.736 & 0.008 & 70.71 & 1.12 & 25 & 50 & - & $-4.425$ & 1481.03 & $-4.197$                      \\
        & 6.4 & 50 & 100 & 0.736 & 0.008 & 70.71 & 4.0 & 1 & - & 300 & $-4.525$ & 706.4 & $-4.674$                          \\

\end{tabular}
\caption{Input parameters and predicted wind properties of all models presented in $\mathrm{Sect.}\,\ref{sec: Hydro_game}$. This includes the so-called `base model' detailed in $\mathrm{Sect.}\,\ref{sec: stellar_param_mass_loss}$, and the different sequences branching from this `base model'.} 
\label{tab: wind_prop_PoWR_HD}
\end{sidewaystable*}

\begin{table*}
\centering
\renewcommand{\arraystretch}{1.15} 
\begin{tabular}{c c c c c} 
        \hline
        & `Base model' (Z = 0.008) & Z = 0.02 & Z = 0.004 & Z = 0.002 \\
        \hline
         $X$ &  0.736 & 0.736 & 0.736 & 0.736 \\
         $Y$ & & $1- X - Z$  \\
         $Z_\mathrm{C}$  & $1.3766\times10^{-3}$ & $3.4416\times10^{-3}$ & $6.8832\times10^{-4}$ & $3.4416\times10^{-4}$\\
         $Z_\mathrm{N}$  & $4.0326\times10^{-4}$ & $1.0082\times10^{-3}$ & $2.0163\times10^{-4}$ & $1.0082\times10^{-4}$\\
         $Z_\mathrm{O}$  & $3.7442\times10^{-3}$ & $9.3605\times10^{-3}$ & $1.8721\times10^{-3}$ & $9.3605\times10^{-4}$\\
         $Z_\mathrm{Ne}$ & $8.3978\times10^{-4}$ & $2.0995\times10^{-3}$ & $4.1989\times10^{-4}$ & $2.0995\times10^{-4}$\\
         $Z_\mathrm{Na}$ & $1.6626\times10^{-5}$ & $4.1565\times10^{-5}$ & $8.3130\times10^{-6}$ & $4.1565\times10^{-6}$\\
         $Z_\mathrm{Mg}$ & $3.1985\times10^{-4}$ & $7.9963\times10^{-4}$ & $1.5993\times10^{-4}$ & $7.9963\times10^{-5}$\\
         $Z_\mathrm{Al}$ & $2.8862\times10^{-5}$ & $7.2154\times10^{-5}$ & $1.4431\times10^{-5}$ & $7.2154\times10^{-6}$\\
         $Z_\mathrm{Si}$ & $3.5297\times10^{-4}$ & $8.8242\times10^{-4}$ & $1.7648\times10^{-4}$ & $8.8242\times10^{-5}$\\
         $Z_\mathrm{P}$  & $3.8927\times10^{-6}$ & $9.7317\times10^{-6}$ & $1.9463\times10^{-6}$ & $9.7317\times10^{-7}$\\
         $Z_\mathrm{S}$  & $1.7591\times10^{-4}$ & $4.3977\times10^{-4}$ & $8.7954\times10^{-5}$ & $4.3977\times10^{-5}$\\
         $Z_\mathrm{Cl}$ & $2.3383\times10^{-6}$ & $5.8458\times10^{-6}$ & $1.1692\times10^{-6}$ & $5.8458\times10^{-7}$\\
         $Z_\mathrm{Ar}$ & $3.4733\times10^{-5}$ & $8.6834\times10^{-5}$ & $1.7367\times10^{-5}$ & $8.6834\times10^{-6}$\\
         $Z_\mathrm{K}$  & $1.8256\times10^{-6}$ & $4.5641\times10^{-6}$ & $9.1281\times10^{-7}$ & $4.5641\times10^{-7}$\\
         $Z_\mathrm{Ca}$ & $3.1057\times10^{-5}$ & $7.7642\times10^{-5}$ & $1.5528\times10^{-5}$ & $7.7642\times10^{-6}$\\
         $Z_\mathrm{Fe}$ & $6.6597\times10^{-4}$ & $1.6649\times10^{-3}$ & $3.3298\times10^{-4}$ & $1.6649\times10^{-4}$\\
        \hline
\end{tabular}
\caption{Mass fractions of different elements used in the `base model' with  total $Z = 0.008$ and the $Z$ model sequence with total $Z$ values of 0.02, 0.004 and 0.002.} 
\label{tab: base_model}
\end{table*}

\begin{table*}
\centering
\renewcommand{\arraystretch}{1.15} 
\begin{tabular}{c c c c} 
        \hline
        & R136a1  & R144-prim & R144-sec \\
        \hline
         $X$ &  0.5 & 0.25 & 0.3 \\
         $Y$ & & $1- X - Z$  \\
         $Z_\mathrm{C}$  & $1.0039\times10^{-4}$ & $7\times10^{-5}$ & $7\times10^{-5}$ \\
         $Z_\mathrm{N}$  & $5.0784\times10^{-3}$ & $4\times10^{-3}$ & $4\times10^{-3}$\\
         $Z_\mathrm{O}$  & $3.4887\times10^{-4}$ &  $1\times10^{-5}$ & $1\times10^{-5}$\\
         $Z_\mathrm{Ne}$ & & $8.3978\times10^{-4}$ \\
         $Z_\mathrm{Na}$ & & $1.6626\times10^{-5}$ \\
         $Z_\mathrm{Mg}$ & & $3.1985\times10^{-4}$ \\
         $Z_\mathrm{Al}$ & & $2.8862\times10^{-5}$ \\
         $Z_\mathrm{Si}$ & & $3.5297\times10^{-4}$ \\
         $Z_\mathrm{P}$  & & $3.8927\times10^{-6}$ \\
         $Z_\mathrm{S}$  & & $1.7591\times10^{-4}$ \\
         $Z_\mathrm{Cl}$ & & $2.3383\times10^{-6}$ \\
         $Z_\mathrm{Ar}$ & & $3.4733\times10^{-5}$ \\
         $Z_\mathrm{K}$  & & $1.8256\times10^{-6}$ \\
         $Z_\mathrm{Ca}$ & & $3.1057\times10^{-5}$ \\
         $Z_\mathrm{Fe}$ & & $6.6597\times10^{-4}$ \\
        \hline
\end{tabular}
\caption{Mass fractions of different elements used in our best-fit $\texttt{PoWR}^\textsc{hd}$ atmosphere models of R136a1 and R144 binary stars.} 
\label{tab: R136_R144_composition}
\end{table*}

\twocolumn
\clearpage

\section{Spectral fits of clumping and $\varv_\mathrm{turb}$ test models}
\label{Appendix: spectra_vturb_clump}

\begin{figure}[h!]
    \includegraphics[width = \columnwidth]{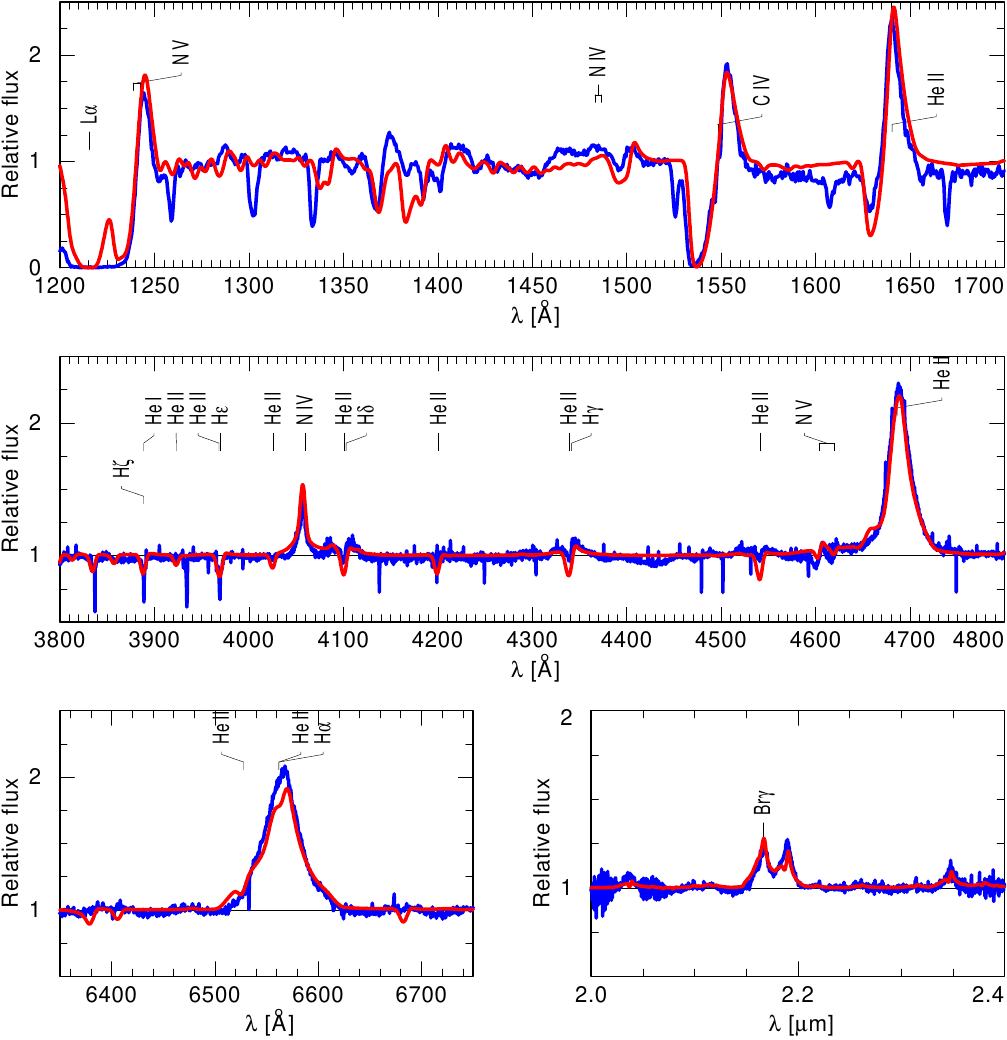}
    \caption{Normalised far-UV HST-STIS/G140L, optical HST-STIS/G430M, H$\alpha$ G750M and K-band VLT/SINFONI observations (blue line), 
    which is compared to the synthetic composite spectra from our best-fit fixed-$M$ R136a1 model with $\varv_\mathrm{turb} = 21.21\,\mathrm{km\,s}^{-1}$ (red line).}
    \label{fig: R136_fixedM}
\end{figure}

\begin{figure}[h!]
    \includegraphics[width = \columnwidth]{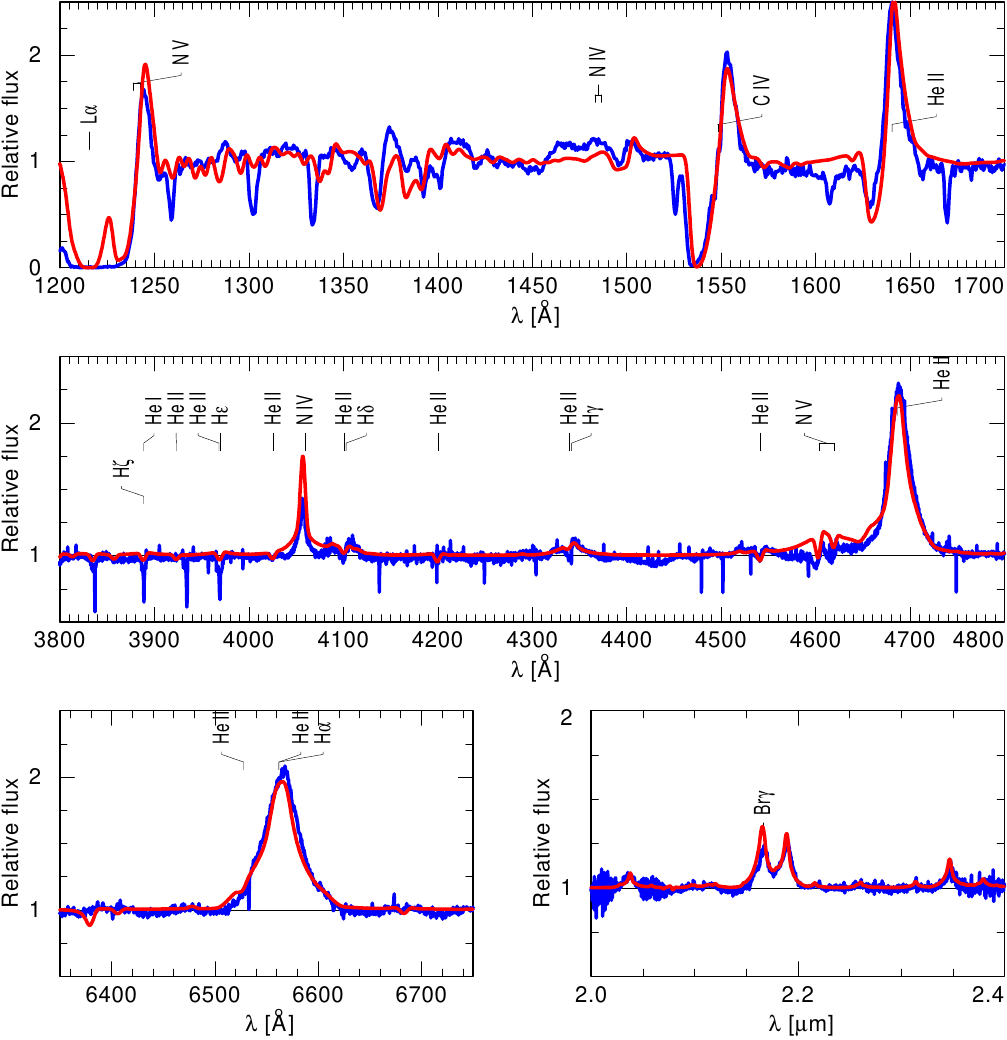}
    \caption{Normalised far-UV HST-STIS/G140L, optical HST-STIS/G430M, H$\alpha$ G750M and K-band VLT/SINFONI observations (blue line), 
    which is compared to the synthetic composite spectra from our best-fit fixed-$M$ R136a1 model with $\varv_\mathrm{turb} = 70.71\,\mathrm{km\,s}^{-1}$ (red line).}
    \label{fig: R136_vturb1}
\end{figure}

\begin{figure}[h!]
    \includegraphics[width = \columnwidth]{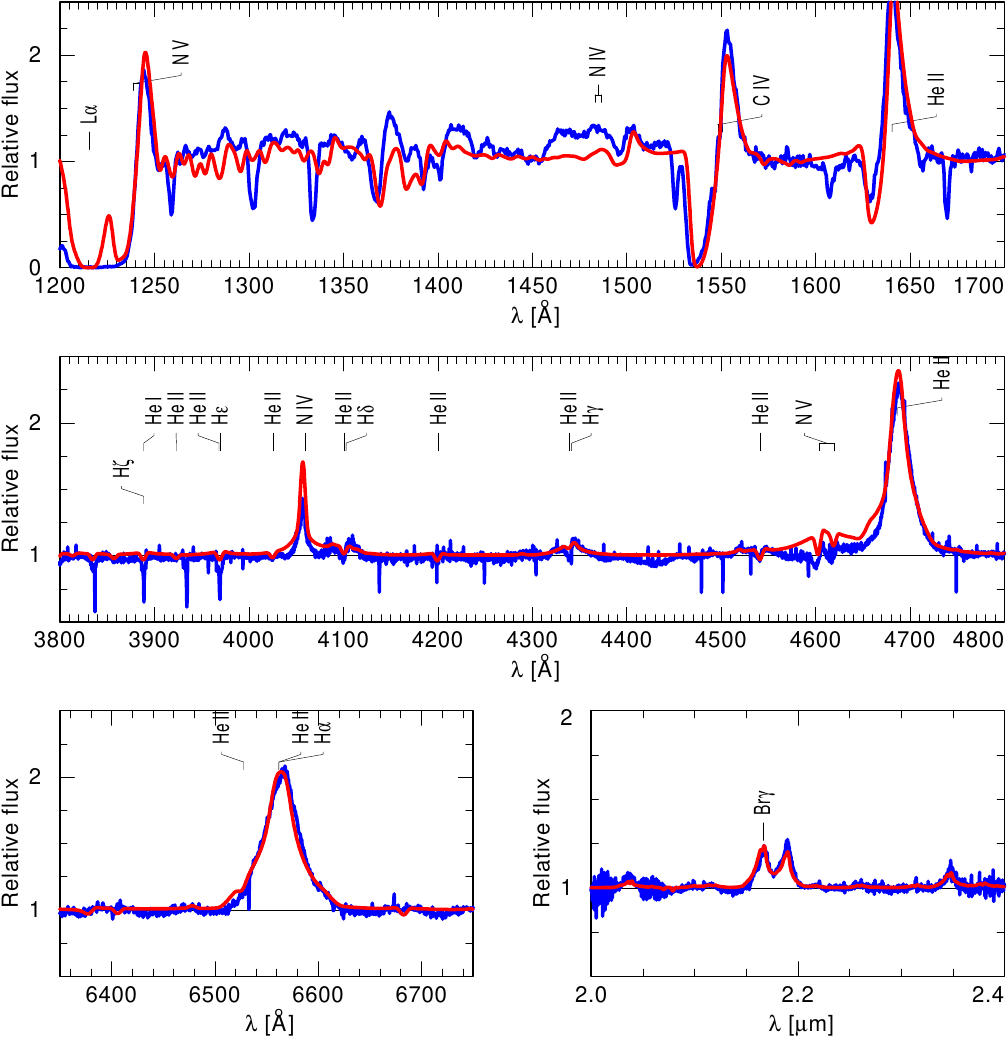}
    \caption{Normalised far-UV HST-STIS/G140L, optical HST-STIS/G430M, H$\alpha$ G750M and K-band VLT/SINFONI observations (blue line), 
    which is compared to the synthetic composite spectra from our best-fit fixed-$M$ R136a1 model with $\varv_\mathrm{turb} = 100\,\mathrm{km\,s}^{-1}$ (red line).}
\end{figure}

\end{document}